\documentclass[lettersize,journal]{IEEEtran}
\usepackage{amsmath,amsfonts}
\usepackage{algorithmic}
\usepackage{algorithm}
\usepackage{array}
\usepackage[caption=false,font=normalsize,labelfont=sf,textfont=sf]{subfig}
\usepackage{textcomp}
\usepackage{stfloats}
\usepackage{url}
\usepackage{verbatim}
\usepackage{graphicx}
\usepackage{cite}
\def\BibTeX{{\rm B\kern-.05em{\sc i\kern-.025em b}\kern-.08em
		T\kern-.1667em\lower.7ex\hbox{E}\kern-.125emX}}
\usepackage{balance}

\usepackage{color, soul}
\usepackage{xcolor}
\soulregister{\cite}7 
\soulregister{\citep}7 
\soulregister{\citet}7 
\soulregister{\ref}7 

\usepackage{tcolorbox}
\usepackage{bm}
\usepackage{bbding}
\usepackage{multirow}
\usepackage{enumitem}
\usepackage{bigstrut}
\usepackage{makecell}
\usepackage[numbers]{natbib}

\usepackage[normalem]{ulem} 

\begin{document}

\title{A Survey of Knowledge Tracing: Models, Variants, and Applications}

\author{ Shuanghong Shen, Qi Liu,~\IEEEmembership{Member,~IEEE}, Zhenya Huang,~\IEEEmembership{Member,~IEEE}, Yonghe Zheng, \\Minghao Yin,~\IEEEmembership{Member,~IEEE}, Minjuan Wang,~\IEEEmembership{Member,~IEEE},  and Enhong Chen,~\IEEEmembership{Fellow,~IEEE}

	\IEEEcompsocitemizethanks{
		\IEEEcompsocthanksitem S.~Shen, and E.~Chen~(corresponding author) are with the School of Data Science, University of Science and Technology of China \& State Key Laboratory of Cognitive Intelligence, Hefei, Anhui, 230026, China.
		\protect Email: closer@mail.ustc.edu.cn, cheneh@ustc.edu.cn
		
	}

	\IEEEcompsocitemizethanks{
	\IEEEcompsocthanksitem Q.~Liu~(corresponding author), and Z.~Huang are with the School of Computer Science and Techonology, University of Science and Technology of China \& State Key Laboratory of Cognitive Intelligence \& Institute of Artificial Intelligence, Hefei Comprehensive National Science Center, Hefei, Anhui, 230026, China.
	\protect Email: \{qiliuql, huangzhy\}@ustc.edu.cn
	
}
	
	\IEEEcompsocitemizethanks{
		\IEEEcompsocthanksitem Y.~Zheng is with the Research Institute of Science Education, Faculty of Education, Beijing Normal University, Haidian, Beijing, 100875, China.
		\protect Email: zhengyonghe@bnu.edu.cn
	
	}

	\IEEEcompsocitemizethanks{
		\IEEEcompsocthanksitem M.~Yin is with the School of Information Science and Technology, Northeast Normal University, Changchun, Jilin, 130117, China.
		\protect Email: ymh@nenu.edu.cn
		
	}
	
	\IEEEcompsocitemizethanks{
		\IEEEcompsocthanksitem M.~Wang is with The Education University of Hong Kong, Hong Kong SAR, 999077, China and San Diego State University, San Diego, CA 92182 USA.
		\protect Email: mwang@sdsu.edu
		
	}

}

\markboth{Journal of \LaTeX\ Class Files,~Vol.~14, No.~8, August~2021}%
{Shell \MakeLowercase{\textit{et al.}}: A Sample Article Using IEEEtran.cls for IEEE Journals}

\maketitle

\begin{abstract}
Modern online education has the capacity to provide intelligent educational services by automatically analyzing substantial amounts of student behavioral data.  Knowledge Tracing (KT) is one of the fundamental tasks for student behavioral data analysis, aiming to monitor students' evolving knowledge state during their problem-solving process. In recent years, a substantial number of studies have concentrated on this rapidly growing field, significantly contributing to its advancements. In this survey, we will conduct a thorough investigation of these progressions. Firstly, we present three types of fundamental KT models with distinct technical routes. Subsequently, we review extensive variants of the fundamental KT models that consider more stringent learning assumptions. Moreover, the development of KT cannot be separated from its applications, thereby we present typical KT applications in various scenarios. To facilitate the work of researchers and practitioners in this field, we have developed two open-source algorithm libraries: EduData that enables the download and preprocessing of KT-related datasets, and EduKTM that provides an extensible and unified implementation of existing mainstream KT models. Finally, we discuss potential directions for future research in this rapidly growing field. We hope that the current survey will assist both researchers and practitioners in fostering the development of KT, thereby benefiting a broader range of students. 
\end{abstract}

\begin{IEEEkeywords}
Online Education, Knowledge Tracing, Educational Data Mining, Adaptive Learning, User Modeling.
\end{IEEEkeywords}

\section{Introduction}
\IEEEPARstart{W}{ith} the proliferation of the Internet and mobile communication technology, online education has become increasingly popular and is now developing at an unprecedented scale \citep{2009Theory,2015Online, wang2021art}.
{This innovative style of learning provides a degree of flexibility that conventional education cannot match, which enables teaching and learning to occur at any time and any place.}
Meanwhile, online learning systems (e.g., Coursera, ASSISTment) \citep{KURT, 8697129, guo2019case} have proven to be even more effective than traditional learning styles, since they can offer more intelligent educational services, such as recommending individualized learning resources to students \citep{jiang2023reinforced, jiang2023courses}. 
To provide these intelligent services, online learning systems continuously record a massive amount of available data about student-system interactions {(e.g., responding to exercises)}, which can be further mined to assess their knowledge levels, learning preferences, and other attributes. Specifically, Knowledge Tracing (KT) \citep{corbett1994knowledge} is one of the most fundamental and critical tasks for analyzing students' learning behavior data, which aims to explore the recorded student-system interactions to monitor their evolving knowledge states \citep{2001Automata,pardos2013adapting}.

\begin{figure*}[t]
	\vspace{-0.5cm}
	\centerline{\includegraphics[width=\textwidth]{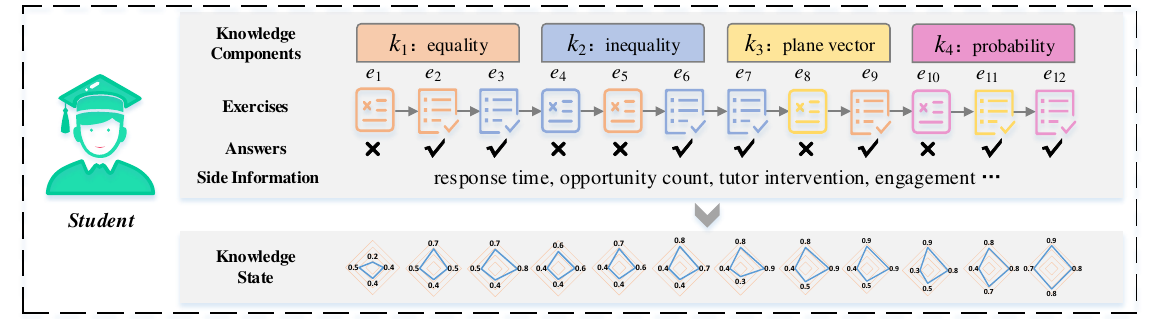}}
	\vspace{-0.2cm}
	\caption{A simple schematic diagram of knowledge tracing. Different knowledge concepts are represented in different colors, while exercises are also depicted in the  color relevant to the knowledge concepts. During the learning process, different kinds of side information are also recorded. The evolving process of the knowledge state is assessed by KT models and illustrated by the radar maps.}
	\label{kt}
	\vspace{-0.5cm}
\end{figure*}

Fig. \ref{kt} presents a simple schematic diagram of knowledge tracing. During the learning process, online learning systems continuously record students' learning behavioral data, including exercises and their related knowledge concepts (e.g., \emph{equality, inequality, plane vector}, \emph{probability}, represented in various colors), and students' answers (i.e., correct or incorrect responses).  A substantial amount of supplementary information is also simultaneously recorded, including response time, opportunity count, and tutor intervention, which provides a more comprehensive reflection of students' learning process.  
Based on the collected learning data, researchers are striving to maintain an estimate of students' evolving knowledge states. {For illustration, we give a case in  Fig. \ref{kt}, where the student's prior knowledge is quantified as 0.2, 0.4, 0.4, and 0.5 across four distinct knowledge concepts. The radar map serves as a visual representation of the student's knowledge mastery, which progressively expands as the student continues to acquire new knowledge in learning.} After a period of learning, the student's knowledge states reach 0.9, 0.8, 0.8, and 0.7 respectively, suggesting good knowledge growth. In the aforementioned learning process, KT models aim to monitor changes in students' knowledge states. Once we understand students' knowledge states, the learning system can customize more suitable learning schemes for different students, thereby enabling the teaching of students in accordance with their proficiency. It also allows students to better comprehend their learning process and gradually focus on improving their skills with poorly mastered concepts  \citep{2013A, liu2019exploiting}.

Knowledge tracing has been studied for decades, with the first studies tracing back to the late 1970s. These initial works primarily focused on confirming the effectiveness of mastery learning \citep{Block1976Mastery}. To the best of our knowledge, ~\citet{corbett1994knowledge} were the first to introduce the concept of knowledge tracing, employing Bayesian networks to model the student learning process, which they referred to as Bayesian Knowledge Tracing. Since then, the significance of KT has been recognized by a broader spectrum of researchers, and increasing attention has been directed towards KT-related research. Many logistic models have been applied to KT, including Learning Factor Analysis \citep{cen2006learning} and Performance Factor Analysis \citep{pavlik2009performance}. In recent years, deep learning has greatly enhanced research into the KT task, largely due to its capacity to extract and represent features and discover intricate structure. For instance, Deep Knowledge Tracing introduced recurrent neural networks (RNNs) \citep{williams1989learning} into the KT task and was found to significantly outperform previous methods \citep{piech2015deep}. Following this, various methods have been introduced that employ various types of neural networks to the KT task, considering various characteristics of the learning sequence \citep{nakagawa2019graph, pandey2019self, CAKT}. Moreover, due to the requirements of practical applications, many variants of KT models have been continuously developed, and KT has already been broadly applied in numerous educational scenarios. 

\begin{figure*}[t]
	\vspace{-0.2cm}
	\centerline{\includegraphics[width=\textwidth]{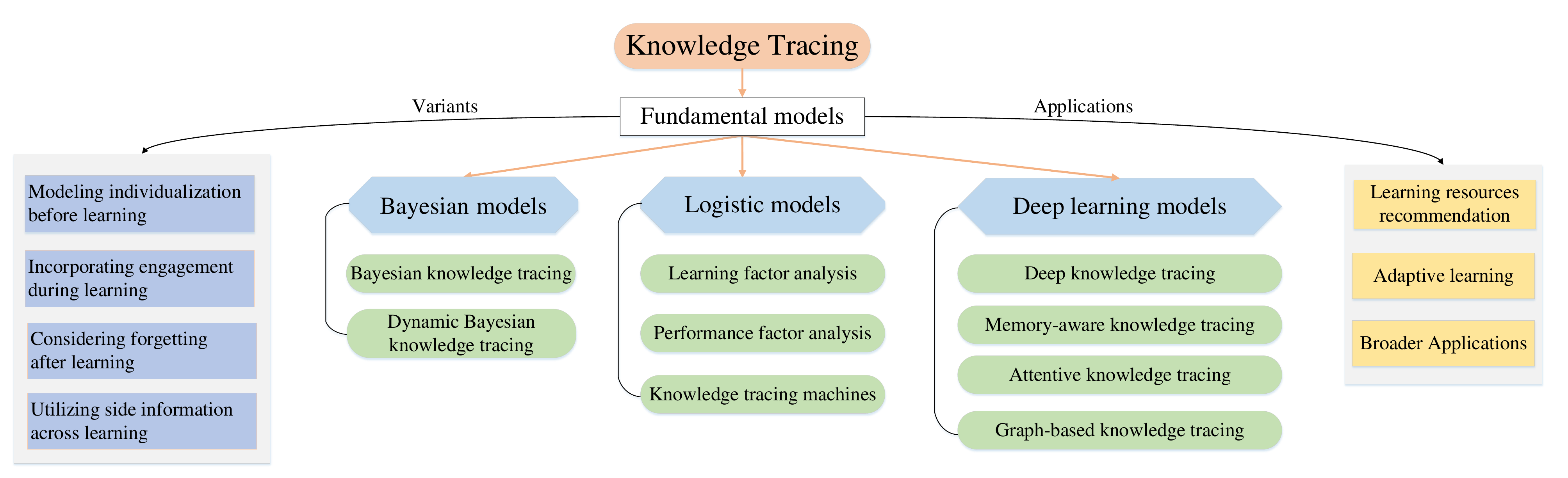}}
	\vspace{-0.2cm}
	\caption{{An overview of knowledge tracing models.} }
	\label{tax}
	\vspace{-0.2cm}
\end{figure*}

While novel KT models continue to emerge, there remains a lack of comprehensive surveys exploring this young research field, particularly {regarding its} numerous variants and applications. To this end, {the current survey aims to systematically review the development of KT}. As depicted in Figure \ref{tax}, we initially categorize existing KT models from a technical perspective, which is consistent with the majority of existing surveys \citep{schmucker2022assessing, abdelrahman2023knowledge}. This categorization splits them into three categories: (1) Bayesian models, (2) logistic models, and (3) deep learning models. In each category, we further organize specific KT methods according to their various techniques. 
Subsequently, we introduce extensive variants of these fundamental KT models, which consider more stringent assumptions about more complete learning process in different learning phases. In addition, we present several typical applications of KT in real learning scenarios. Due to the complexity of different KT models, we have open sourced two algorithm libraries to better aid researchers and practitioners in implementing KT models and facilitate community development in this domain. These libraries, EduData\footnote{https://github.com/bigdata-ustc/EduData} and EduKTM\footnote{https://github.com/bigdata-ustc/EduKTM}, include most existing KT-related datasets, extensible, and unified implementations of existing KT models, and relevant resources. Finally, we discuss potential future research directions. In summary, this paper presents an extensive survey of KT that can serve as a comprehensive guide for both researchers and practitioners. 

The remainder of this survey is structured as follows. Section \ref{sec:overview} presents an overview of the KT task and we discuss the differences between this and previous surveys. Section \ref{sec:models} provides a review of the three categories of fundamental KT models. 
Section \ref{sec:variants} describes the variants of fundamental KT models. 
Section \ref{sec:applications} introduces the extensive applications of KT in different scenarios. Section \ref{sec:dataset} gives the summary of existing datasets for evaluating KT models and details of the algorithm libraries we have released. Section \ref{sec:future} discusses some potential future research directions. Finally, section \ref{sec:conclution} summarizes the paper.

\section{Overview} \label{sec:overview}

\subsection{Problem Definition}

In an online learning system, supposing there exists a set of students $\mathbb{S}$ and a set of exercises $\mathbb{E}$. {Each exercise is related to specific Knowledge Concepts (KCs). Generally, the name given to the knowledge related to exercises differs across online learning platforms. For instance, it is named \emph{skill} in ASSISTments \citep{4967564}. To promote better understanding, we refer to these uniformly as knowledge concepts throughout this paper, and denote the set of all KCs as $\mathbb{KC}$. Moreover,  $M$ and $K$ are respectively used to represent the total number of exercises and KCs. Students are asked to answer different exercises in order to achieve mastery of the related knowledge.}
Therefore, the learning sequence of a student can be formulated as $\bm{X}  = \{([e_1, k_{e_1}], a_1, r_1), ([e_2, k_{e_2}], a_2, r_2), ..., ([e_t, k_{e_t}], a_t, r_t), ..., \\([e_N, k_{e_N}], a_N, r_N)\}$, where the tuple $([e_t, k_{e_t}], a_t, r_t)$ represents the learning interaction at the $t-$th time step, $e_t$ represents the exercise,  $k_{e_t}$ represents the exercise's related KCs,  $a_t$ represents the correctness label (i.e., with 1 for correct and 0 for incorrect answers),  $r_t$ stands for the side information recorded in this learning interaction, and $N$ is the length of the learning sequence. The research problem of knowledge tracing can thus be defined as follows:

\textbf{Given sequences of learning interactions in online learning systems, knowledge tracing aims to monitor students' evolving knowledge states during the learning process and predict their performance on future exercises. The measured knowledge states can be further applied to individualize students' learning schemes in order to maximize their learning efficiency.}

Some recent works directly regarded the KT task as student performance prediction, without considering students' knowledge states \citep{vaswani2017attention, SKVMN_plus, SAINT_plus}. We agree that predicting student performance is of great significance, as it is now {the best way} to evaluate the quality of the knowledge state traced by KT models.  However, we have to point out that KT focuses more on students' knowledge states, especially their interpretability and rationality, which is related to the students' acceptance of the conclusions given based on the KT model \citep{9950313, 10093075}.

\subsection{Categorization}
{ As shown in Fig. \ref{tax}, we categorize and summarize the existing KT models according to their technical differences. In more detail, the proposed taxonomy splits existing KT methods into three categories: (1) Bayesian models, which are implemented through probability model, (2) logistic models,  which are implemented through logistic functions, and (3) deep learning models, which are implemented through neural networks. Specifically, we divide the deep learning models into four sub-categories according to four various neural networks, i.e., deep knowledge tracing based on recurrent neural networks, memory-aware knowledge tracing based on memory networks, attentive knowledge tracing based on self-attention mechanism, and graph-based knowledge tracing based on graph neural networks.
In addition to these fundamental KT models, we also introduce a large number of their variants, which respectively consider a more complete learning process in distinct learning phases: (1) modeling individualization before learning, (2) incorporating engagement during learning, (3) considering forgetting after learning, and (4) utilizing side information across learning. Moreover, we also summarize the extensive applications of KT in different educational scenarios, including learning resources recommendation, adaptive learning and broader applications beyond student learning.}

\subsection{Differences between this and previous surveys}

Given the increasing importance of KT, several recent surveys have also examined this area. These include works by \citet{pelanek2017bayesian, schmucker2022assessing, song2022survey, liu2022knowledge, abdelrahman2023knowledge}, \citet{zanellati2024hybrid}. Here, we will briefly discuss the key distinctions between these studies to highlight the necessity and significance of this survey. 

Existing surveys have either focused on specific categories of KT models or comprehensively reviewed all available KT models.  {For example, \citet{pelanek2017bayesian} provided an overview of KT in terms of Bayesian models and logistic models, \citet{song2022survey} compared and discussed deep learning based KT models, while \citet{zanellati2024hybrid} presented to pay more attention to hybird models in KT.} \citet{liu2022knowledge} conducted a bibliometric analysis to examine the evolution of KT research from 1992 to 2021. \citet{abdelrahman2023knowledge} also presented a comprehensive survey for the KT literature, including a broad range of methods starting from the early attempts to the recent state-of-the-art techniques utilizing deep learning. \citet{schmucker2022assessing} summarized KT methods in the context of student performance modeling problems. 

However, current surveys are somewhat limited in their scope, they only provide a detailed introduction to various KT methods and comparisons between them. Given the complexity of online learning systems and the significant importance of KT research in practical applications, this survey places a greater emphasis on the variants and applications of KT models, rather than solely introducing and comparing different KT methods. Moreover, considering that datasets are collected from different systems with various setting, subjects, learning stages, and scales,  
we do not report and compare the performance of KT models on the student performance prediction task across various datasets in this survey.  \citet{pavlik2021logistic} have also empirically verified that no single KT model was always the best, a specific better model must consider multiple student features and the learning context.
Instead, we have open sourced two algorithm libraries which include the majority of  existing KT-related datasets and unified implementations of existing KT models. Consequently, researchers and practitioners can freely select appropriate KT models based on their specific requirements in various application scenarios. 

\begin{table*}[t]
	\vspace{-0.6cm}
	\centering
	\renewcommand\arraystretch{1.0}
	\caption{A summary of different types of fundamental knowledge tracing models.}
	\vspace{-0.3cm}
	\resizebox{\textwidth}{!}{
		\begin{tabular}{|c|c|c|c|c|}
			\hline
			\multicolumn{1}{|c|}{Category} & Typical approach & Technique & KC relationship & Knowledge state \bigstrut\\
			\hline
			\multicolumn{1}{|c|}{\multirow{2}[5]{*}{Bayesian models}} & Bayesian knowledge tracing \citep{corbett1994knowledge} & Bayesian networks & independent & \multirow{2}[5]{*}{\shortstack{unobservable node \\ in HMM}} \bigstrut\\
			\cline{2-4}          & dynamic Bayesian knowledge tracing \citep{kaser2017dynamic} & dynamic Bayesian networks & pre-defined & \multicolumn{1}{c|}{} \bigstrut\\
			\hline
			\multicolumn{1}{|c|}{\multirow{3}[8]{*}{Logistic models}} & learning factor analysis \citep{cen2006learning} & \multirow{2}[4]{*}{logistic regression} & \multirow{3}[8]{*}{independent} & \multirow{3}[6]{*}{\shortstack{the output of \\ logistic regression function}} \bigstrut\\
			\cline{2-2}          & performance factor analysis \citep{pavlik2009performance} & \multicolumn{1}{c|}{} & \multicolumn{1}{c|}{} & \multicolumn{1}{c|}{} \bigstrut\\
			\cline{2-3}          & knowledge tracing machines \citep{vie2019knowledge}& factorization machines & \multicolumn{1}{c|}{} & \multicolumn{1}{c|}{} \bigstrut\\
			\hline
			\multicolumn{1}{|c|}{\multirow{4}[6]{*}{\shortstack{Deep learning \\ models}}} & deep knowledge tracing \citep{piech2015deep} & RNN/LSTM & discover automatically & the hidden state \bigstrut\\
			\cline{2-5}          & memory-aware knowledge tracing \citep{zhang2017dynamic} & memory networks & correlation weights & \emph{value} matrix \bigstrut\\
			\cline{2-5}          & attentive knowledge tracing \citep{pandey2019self,CAKT} & self-attention mechanism & attention weights &  attentive historical knowledge state \\
			\cline{2-5}          & graph-based knowledge tracing  \citep{nakagawa2019graph} & graph neural networks & edges in graph & aggregate in the graph \bigstrut\\
			\hline
		\end{tabular}
	}
	\vspace{-0.6cm}
	\label{tab1}%
\end{table*}%

\section{Fundamental Knowledge Tracing Models} \label{sec:models}

In this section, as shown in Table \ref{tab1}, we will present the fundamental KT models, based on our taxonomic framework. {Specifically, we will introduce these models in accordance with their development timeline. Subsequently, we will provide a summary of these fundamental KT models.}

\subsection{Bayesian Models}
Bayesian models assume that the learning process adheres to a Markov process. This process allows for the estimation of students' latent knowledge states based on their observed performance. In the subsequent section, we will present two Bayesian models in our taxonomy framework: the Bayesian Knowledge Tracing (BKT) and the Dynamic Bayesian Knowledge Tracing (DBKT). 

\subsubsection{Bayesian Knowledge Tracing}
\begin{figure}[t]
	\vspace{-0.0cm}
	\centerline{\includegraphics[width=0.9\columnwidth]{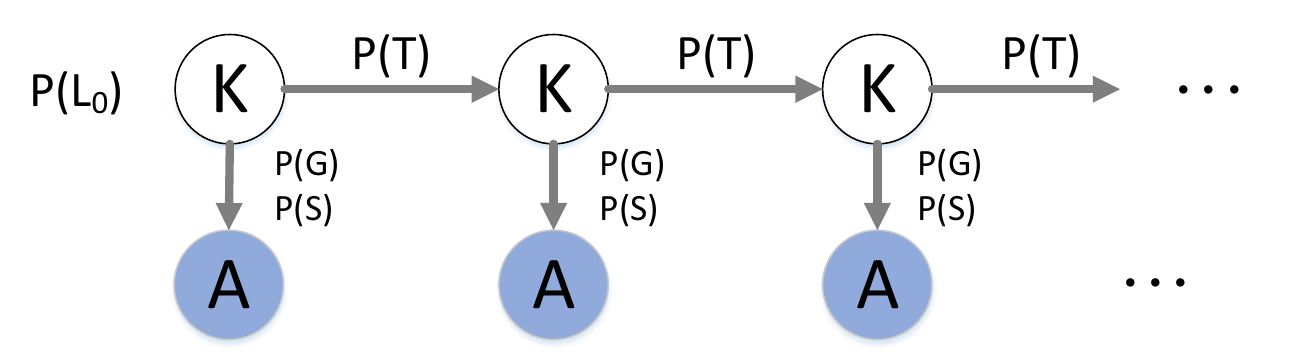}}
	\vspace{-0.2cm}
	\caption{The topology of Bayesian Knowledge Tracing \citep{corbett1994knowledge}. $K$ are the unobserved knowledge nodes, $A$ are the observed performance (answer) nodes, $P(L_0)$ represents the initial probability, $P(T)$ is the transition probability, $P(G)$ is the guessing probability  and $P(S)$ is the slipping probability.}
	\label{fbkt}
	\vspace{-0.3cm}
\end{figure}

{BKT's structure is illustrated in Fig. \ref{fbkt}; here, the unshaded nodes represent unobservable latent knowledge states, while the shaded nodes represent the observable answers of the student.}

BKT is a unique instance of the Hidden Markov Model (HMM). There are two types of parameters in HMM: transition probabilities and emission probabilities. In BKT, the transition probabilities are defined by two learning parameters: (1)  $P(T)$, the probability of transition from the unlearned state to the learned state; (2) $P(F)$, the probability of forgetting previously mastered knowledge. Moreover, the emission probabilities are determined by two performance parameters: (1) $P(G)$ - the probability that a student will guess correctly, despite non-mastery; (2) $P(S)$ - the probability that a student will make a mistake, despite mastery. Additionally, $P(L_0)$ represents the initial probability of mastery. BKT operates within a two-state student modeling framework: knowledge is either learned or unlearned, and there is no forgetting once a student has mastered the knowledge. Based on the observations of the student's learning interactions, the following equation is utilized to estimate the knowledge state and the probability of correct answers: 
\begin{equation}
	\small
	\begin{aligned}
		P(L_n) &= P(L_{n}|Answer) + (1 -  P(L_{n}|Answer))P(T), \\
		P(C_{n+1}) &= P(L_n)(1 - P(S)) + (1 - P(L_n))P(G),
	\end{aligned}
\end{equation}
where $P(L_n)$ is the probability that a KC is mastered at the $n$-th learning interaction, $P(C_{n+1})$ is the probability of correct answers at the next learning interaction. $P(L_n)$ is the sum of two probabilities: (1) the probability that the KC is already mastered; (2) the probability that the knowledge state will convert to the mastered state. The posterior probability $P(L_{n}|Answer)$ is estimated as follows:
\begin{equation}
	\scriptsize
	\begin{aligned}
		P(L_{n}|correct)&=
		\frac{P(L_{n-1})(1 - P(S))}{P(L_{n-1})(1 - P(S)) + (1 - P(L_{n-1}))P(G)}, \\
		P(L_{n}|incorrect)&=
		\frac{P(L_{n-1})P(S)}{P(L_{n-1})P(S) + (1 - P(L_{n-1}))(1 - P(G))}.
	\end{aligned}
\end{equation}

\subsubsection{Dynamic Bayesian Knowledge Tracing} \label{dynamic}
BKT independently models the parameters of each KC, employing a specific model for each KC. However, KCs are not completely independent, but rather hierarchical and closely related \citep{huang2019constructing}. Dynamic Bayesian networks are capable of jointly representing multiple skills within a single model, potentially enhancing the representational power of BKT. Consequently, \citet{kaser2017dynamic} propose dynamic Bayesian knowledge tracing (DBKT) to represent the hierarchies and relationships within KCs using dynamic Bayesian networks. This approach considers different KCs jointly within a single model. 

In DBKT, a student's knowledge mastery is represented by binary latent variables, which are estimated based on their learning interactions. Unlike BKT, DBKT takes into account the dependencies between various KCs. For example, if $KC_1$ and $KC_2$ are prerequisites for mastering $KC_3$, students' mastery of $KC_3$ depends on their mastery of $KC_1$ and $KC_2$.
Let $H$ denote the unobserved variables, i.e., lack of student answers and binary mastery variables. Assuming the student correctly answers an exercise associated with $KC_1$ at time step $t_1$, i.e., $a_{1,1} = 1$. The observed variables are then $a_m = a_{1,1}$ and the unobserved variables are $h_m = \{KC_{1,1}, KC_{2,1}, KC_{3,1}, a_{2,1}, a_{3,1}\}$. The objective of DBKT is to find the parameters $\theta$ that maximize the joint probability $p(a_m, h_m|\theta)$. The log-likelihood can alternatively be formulated using a log-linear model, as follows: 
\begin{equation}
	L(\bm{w}) = \sum_{m}ln(\sum_{h_m}exp(\bm{w}^T\varPhi(a_m, h_m) - ln(Z))),
	\vspace{-0.2cm}
\end{equation}
where $\varPhi: A \times H \rightarrow \mathbb{R}^{F}$ denotes a mapping from the observed space $A$ and the latent space $H$ to an $F$-dimensional feature vector. $Z$ is a normalizing constant, $\bm{w}$ denotes the weights.

\subsection{Logistic Models}

{Logistic models represent the probability of students correctly answering exercises as a logistic function of the student and KC parameters.} They first use different factors in students' learning interactions to compute an estimation of the student and KC parameters, then utilize a logistic function to transform this estimation into the prediction of the probability of mastery \citep{pelanek2017bayesian}.
In the subsequent section, we will introduce three types of logistic models: (1) Learning Factor Analysis (LFA), (2) Performance Factor Analysis (PFA) and (3) Knowledge Tracing Machines (KTM).

\subsubsection{Learning Factor Analysis}
The LFA model \citep{cen2006learning} considers the following learning factors:
\begin{itemize}
	\item{Initial knowledge state}: parameter $\alpha$ estimates the initial knowledge state of each student;
	\item{Easiness of KCs}: parameter $\beta$ captures the easiness of different KCs;
	\item{Learning rate of KCs}: parameter $\gamma$ denotes the learning rate of KCs.
\end{itemize}
The standard LFA model takes the following form:
\vspace{-0.2cm}
\begin{equation}
	p(\theta) = \sigma(\sum_{i \in N}\alpha_iS_i	+ \sum_{j \in KCs}(\beta_j + \gamma_jT_j)K_j), 
	\vspace{-0.2cm}
\end{equation}
where $\sigma$ is the sigmoid function, $S_i$ is the covariate for the student $i$, $T_j$ represents the covariate for the number of interactions on KC $j$, $K_j$ is the covariate for KC $j$, $p(\theta)$ is the estimation of the probability of a correct answer.

\subsubsection{Performance Factor Analysis}
The PFA model \citep{pavlik2009performance} can be seen as an extension of the LFA model that is especially sensitive to the student performance. In contrast to the LFA model, PFA  considers the following different factors:
\begin{itemize}
	\item{Previous failures}: parameter $f$ is the prior failures for the KC of the student;
	\item{Previous successes}: parameter $s$ represents the prior successes for the KC of the student;
	\item{Easiness of KCs}: parameter $\beta$ means the easiness of different KCs, which is the same as in the LFA model.
\end{itemize}
The standard PFA model takes the following form:
\vspace{-0.15cm}
\begin{equation}
	p(\theta) = \sigma(\sum_{j \in KCs}(\beta_j + \mu_js_{ij} + \nu_jf_{ij})),
	\label{pfa}
	\vspace{-0.15cm}
\end{equation}
where $\mu$ and $\nu$ are the coefficients for $s$ and $f$, which denote the learning rates
for successes and failures.

\begin{figure}[t]
	\vspace{-0.2cm}
	\centerline{\includegraphics[width= 0.9\columnwidth]{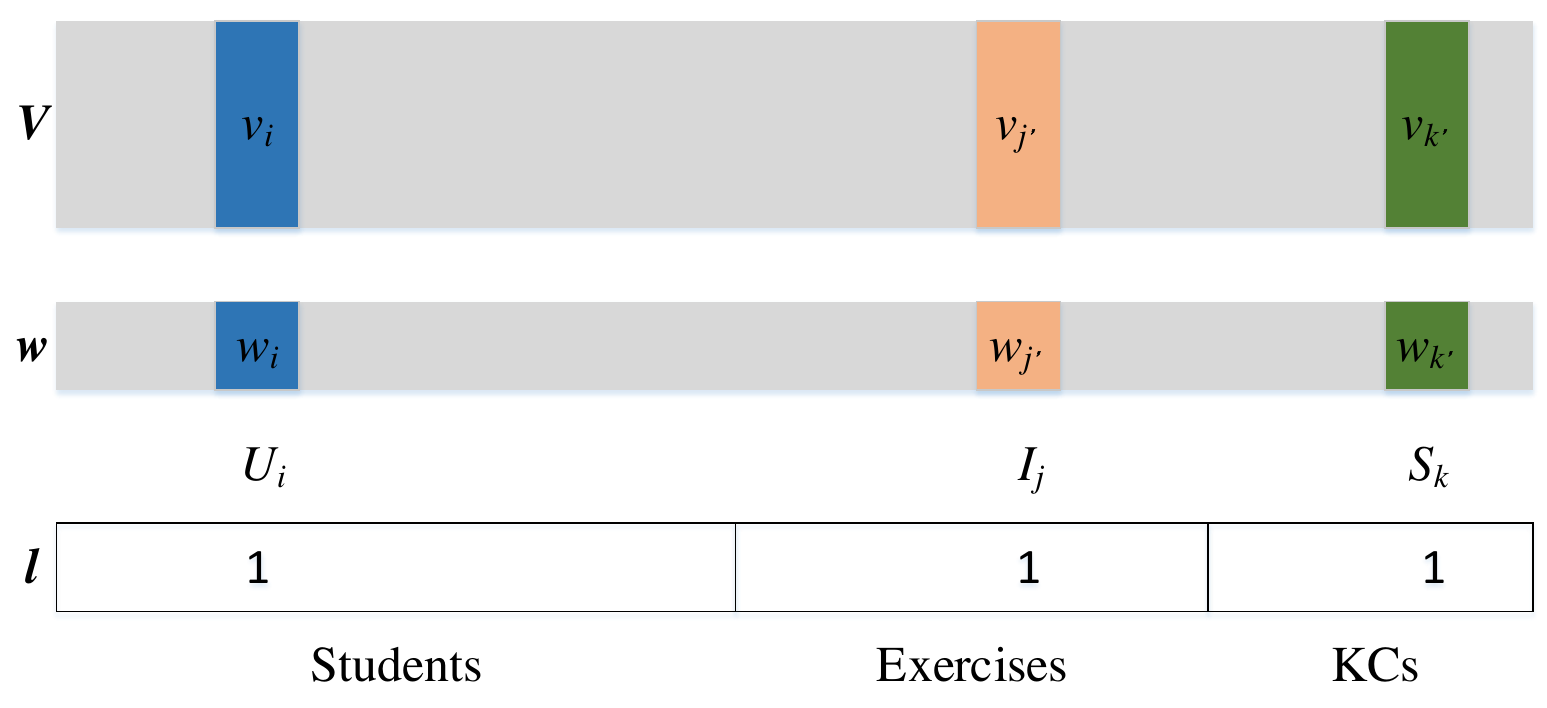}}
	\vspace{-0.2cm}
	\caption{Example of activation of a knowledge tracing machine \citep{vie2019knowledge}. $V$ refers to the matrix of embeddings, $w$ refers to the vector of biases, $x$ is the encoding vector of the learning interaction.}
	\label{fktm}
	\vspace{-0.6cm}
\end{figure}

\subsubsection{Knowledge Tracing Machines}
The KTM model, developed by Vie et al. \citep{vie2019knowledge}, employs factorization machines (FMs) \citep{thai2012factorization, thai2012using} to generalize logistic models to higher dimensions. FMs were initially introduced as a general predictor capable of working with any real-valued feature vector, enabling the model to represent all interactions between variables using factorized parameters \citep{5694074}. FMs provide a means of encoding side information about exercises or students into the model. Figure \ref{fktm} illustrates the example of KTM, which models the knowledge mastery of the student based on a sparse set of weights for all features involved in the event. Let $L$ be the number of features; here, the features can be related to students, exercises, KCs, or any other side information. The learning interaction is encoded by a sparse vector $\bm{l}$ of length $L$. When feature $i$ is involved in the interaction, $l_i > 0$. The probability $p(\theta)$ of the correct answer is determined by the following equations: 
\vspace{-0.3cm}
\begin{equation}
	p(\theta) = \sigma(\mu + \sum_{i = 1}^{L}w_il_i + \sum_{1 \leq i < j \leq L}l_il_j\langle\bm{v_i}, \bm{v_j}\rangle ),
	\vspace{-0.2cm}
\end{equation}
where $\mu$ is the global bias, the feature $i$ is modeled by the bias $w_i \in \bm{R}$ and the embedding $\bm{v}_i \in \bm{R}^d$ ($d$ is the dimension). Note that only features with $l_i > 0$ will have impacts on the predictions.

\subsection{Deep Learning Models} \label{deep}
The cognitive process can be influenced by various factors at both the macro and micro levels. It is difficult for Bayesian models or logistic models to adequately capture a cognitive process of high complexity \citep{piech2015deep}.
Deep learning, with its potent ability to achieve non-linearity and feature extraction, is well-suited for modeling complex learning processes, particularly when a significant amount of learning interaction data is available \citep{khajah2016deep}. {In recent years, numerous research works have been proposed on deep learning KT models, we will introduce deep learning models from four sub-categories: (1) deep knowledge tracing, (2) memory-aware knowledge tracing, (3) attentive knowledge tracing, and (4) graph-based knowledge tracing.}

\subsubsection{Deep Knowledge Tracing}

Deep Knowledge Tracing (DKT) is the pioneering approach that introduces deep learning {to complete the KT task. DKT employs Recurrent Neural Networks (RNNs) \citep{williams1989learning} to process the input sequence of learning interactions over time,} maintaining a hidden state that implicitly contains information about the history of all past elements of the sequence. This hidden state evolves based on both the previous knowledge state and the present input learning interaction \citep{piech2015deep}. DKT provides a high-dimensional and continuous representation of the knowledge state, enabling it to more effectively model the complex learning process. Typically, RNNs' variant, the Long Short-Term Memory (LSTM) networks \citep{hochreiter1997long}, are more frequently used in the implementation of DKT, which is further strengthened by considering forgetting. 

\begin{figure}[t]
	\vspace{-0.3cm}
	\centerline{\includegraphics[width=0.9\columnwidth]{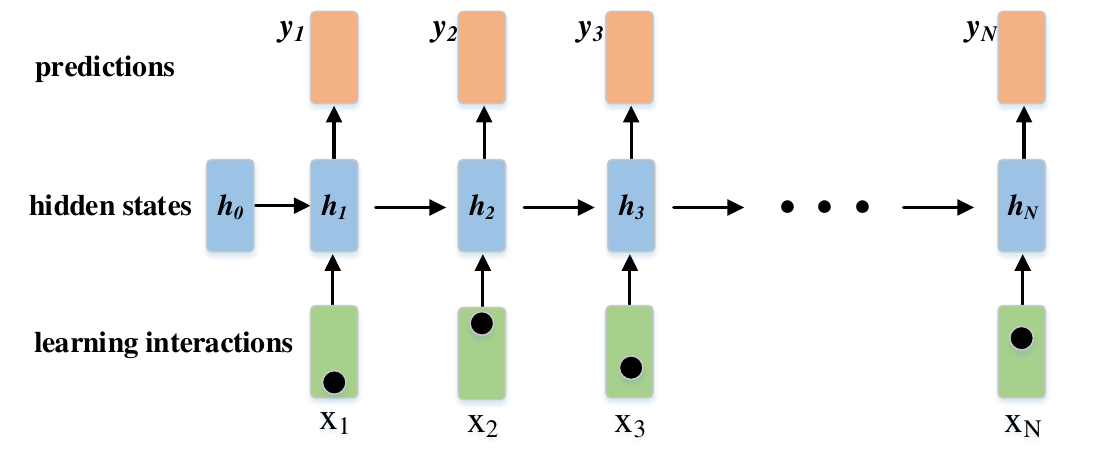}}
	\vspace{-0.3cm}
	\caption{The architecture of DKT \citep{piech2015deep}.}
	\label{fdkt}
	\vspace{-0.6cm}
\end{figure}

Fig. \ref{fdkt} illustrates the process of deep knowledge tracing. In DKT, exercises are represented by their contained KCs. For datasets with different numbers of KCs, DKT applies two different methods to convert students' learning interactions  $\bm{X} = \{(e_1, a_1), (e_2, a_2), ..., (e_t, a_t), ..., (e_N, a_N)\}$ into a sequence of fixed-length input vectors. More specifically, for datasets with a small number $K$ of unique KCs, $\bm{x}_t \in \{0,1\}^{2K}$ is set as a one-hot embedding, where $ \bm{x}_t^k = 1 $ if the answer $a_t$ of the exercise with KC $k$ was correct or $\bm{x}_t^{k + K} = 1$ if the answer was incorrect. For datasets with a large number of unique KCs, one-hot embeddings are considered too sparse. Therefore, DKT assigns each input vector $\bm{x}_t$ to a corresponding random vector, and then uses the embedded learning sequence as the input of RNNs. A linear mapping and activation function are then applied to the output hidden states to obtain the knowledge state of students:
\vspace{-0.1cm}
\begin{equation}
	\begin{aligned}\label{dkt}
		&\bm{h}_t = tanh(\bm{W}_{hs}\bm{x}_t + \bm{W}_{hh}\bm{h}_{t-1} + \bm{b}_h), \\
		&\bm{y}_t = \sigma (\bm{W}_{yh}\bm{h}_{t} + \bm{b}_y),
	\end{aligned}
	\vspace{-0.1cm}
\end{equation}
where $tanh$ is the activation function, $\bm{W}_{hs}$ is the input weights, $\bm{W}_{hh}$ is the recurrent weights, $\bm{W}_{yh}$ is the readout weights, and $\bm{b}_h$ and $\bm{b}_y$ are the bias terms.

Despite demonstrating superior performance compared to Bayesian and logistic models, DKT has several inherent shortcomings. For instance, the lack of interpretability is a significant drawback. It is challenging to understand how the hidden states represent students' knowledge states, and the model cannot explicitly determine a student's knowledge mastery from the hidden state \citep{khajah2016deep}. Additionally, \citet{yeung2018addressing} identified two unreasonable phenomena in DKT that contravene common sense. These are: (1) the inability to reconstruct observed input, and (2) inconsistent predicted knowledge states across time-steps. However, despite these shortcomings, DKT remains a promising KT model \citep{xiong2016going}.

\subsubsection{Memory-aware Knowledge Tracing}
To enhance the interpretability of DKT, memory-aware knowledge tracing introduces an external memory module, as proposed by \citep{graves2016hybrid}. This module is designed to {store the and update the corresponding knowledge mastery of the student.} The most representative example is Dynamic Key-Value Memory Networks (DKVMN) for knowledge tracing, as proposed by \citep{zhang2017dynamic}. DKVMN highlights students' specific knowledge states on various knowledge categories. It initializes a static matrix, referred to as a $key$ matrix to store latent KCs and a dynamic matrix, called a $value$ matrix to store and update the mastery of corresponding KCs through read and write operations over time.

\begin{figure}[t]

	\centerline{\includegraphics[width=0.9\columnwidth]{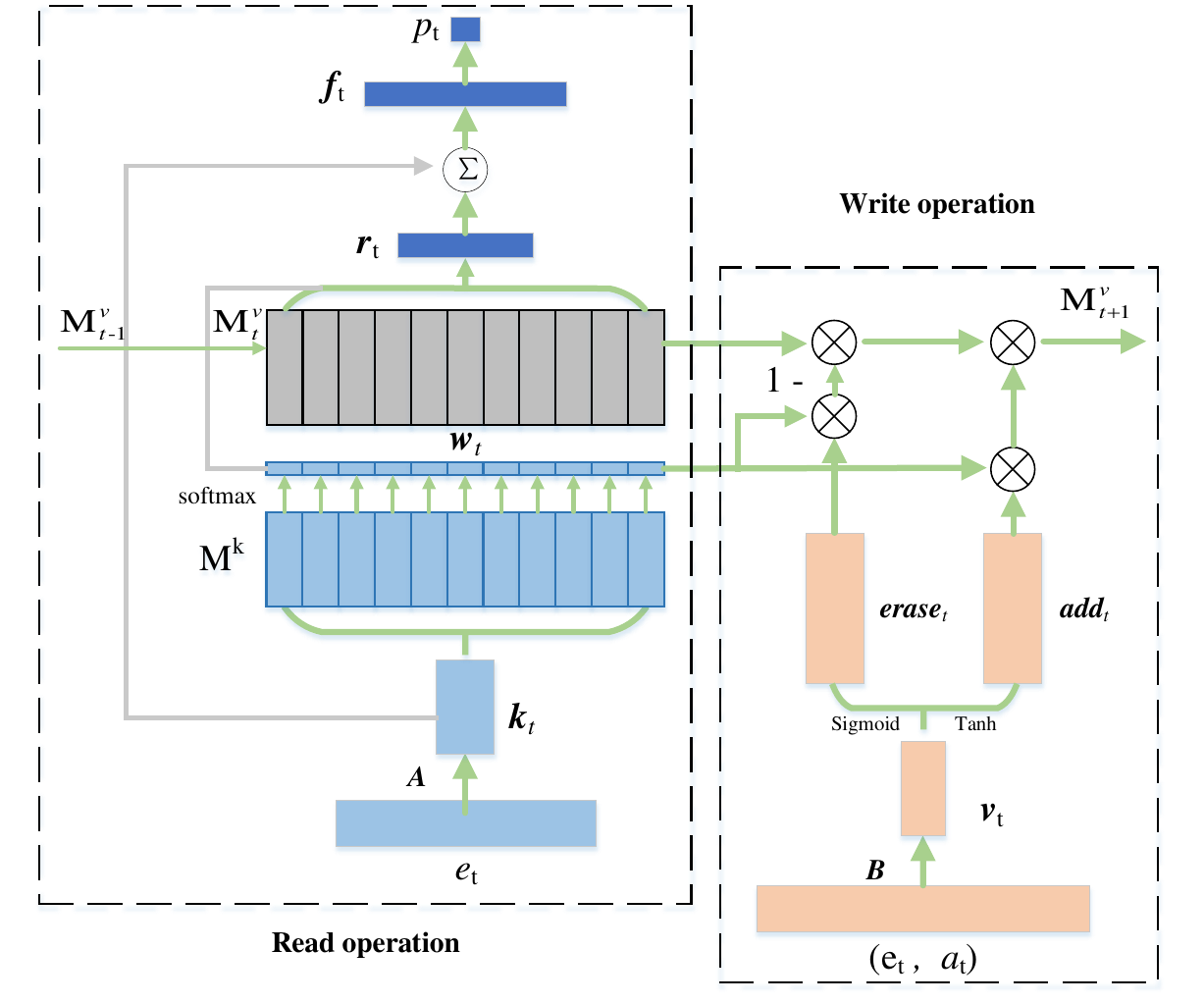}}
	\vspace{-0.3cm}
	\caption{The architecture of DKVMN \citep{zhang2017dynamic}.}
	\label{fdkvmn}
	\vspace{-0.6cm}
\end{figure}

As shown in Fig. \ref{fdkvmn}, an embedding matrix is first defined to obtain the embedding vector $k_t$ of the exercises. A correlation weight $\bm{w}_t$ is then obtained by taking the inner product between the exercise embedding $k_t$ and the $key$ vectors $M^k$, followed by the softmax activation:
\vspace{-0.2cm}
\begin{equation}
	\bm{w}_t = Softmax(k_tM^k),
	\vspace{-0.2cm}
\end{equation}
where the correlation weight $\bm{w}_t$ represents the correlation between the exercises and all latent KCs.

In the read operation, DKVMN predicts student performance based on the student's knowledge mastery. Specifically, DKVMN reads students’ mastery of the exercise $\bm{r}_t$ with reference to the weighted sum of all memory vectors in the $value$ matrix using the correlation weight. The read content and the input exercise embeddings are then concatenated together and passed to a fully connected layer to yield a summary vector $\bm{f}_t$, which contains both the student's knowledge mastery and the prior difficulty of the exercise. Furthermore, the student's performance can be predicted by applying another fully connected layer with a sigmoid activation function to the summary vector:
\vspace{-0.2cm}
\begin{equation}
	\begin{aligned}
		&\bm{r}_t = \sum_{i=1}^{N}w_t(i)M_t^v(i), \\
		&\bm{f}_t = tanh(\bm{W}_f[\bm{r}_t, k_t] + \bm{b}_f),\\
		&p_t = \sigma(\bm{W}_p\bm{f}_t + \bm{b}_p),
	\end{aligned}
	\vspace{-0.1cm}
\end{equation}
where $\bm{W}_f$ and $\bm{W}_p$ are the weights, $\bm{b}_f$ and $\bm{b}_p$ are bias terms.

In the write operation, after an exercise has been answered, DKVMN updates students' knowledge mastery (i.e., the $value$ matrix) based on their performance. Specifically, the learning interaction $(e_t, a_t)$ is first embedded with an embedding matrix $\bm{B}$ to obtain the student's knowledge growth $\bm{v}_t$. Then DKVMN calculates an erase vector $\bm{erase}_t$ from $\bm{v}_t$ and decides to erase the previous memory with reference to both the erase vector and the correlation weight $\bm{w}_t$.
Following erasure, the new memory vectors are updated by the new knowledge state and the add vector $\bm{add}_t$, which forms an $erase$-followed-by-$add$ mechanism that allows forgetting and strengthening knowledge mastery in the learning process:
\vspace{-0.1cm}
\begin{equation}
	\begin{aligned}
		&\bm{erase}_t = \sigma(\bm{W}_e\bm{v}_t + \bm{b}_e), \\
		&\widetilde{M}_t^v(i) = M_{t-1}^v(i)[1 - w_t(i)\bm{erase}_t],\\
		&\bm{add}_t = tanh(\bm{W}_d\bm{v}_t + \bm{b}_d),\\
		&M_{t}^v(i) = \widetilde{M}_t^v(i) + w_t(i)\bm{add}_t,\\
	\end{aligned}
	\vspace{-0.1cm}
\end{equation}
where $\bm{W}_e$ and $\bm{W}_d$ are the weights,  $\bm{b}_e$ and $\bm{b}_d$ are bias terms.

\citet{SKVMN} point out that DKVMN failed to capture long-term dependencies in the learning process. Therefore, they propose a Sequential Key-Value Memory Network (SKVMN) to combine the strengths of DKT's recurrent modelling capacity and DKVMN's memory capacity. In SKVMN, a modified LSTM called \emph{Hop-LSTM} is used to hop across LSTM cells according to the relevance of the latent KCs, which directly captures the long-term dependencies. During the writing process, SKVMN allows for the calculation of the knowledge growth of a new exercise, taking into consideration the current knowledge state, thereby yielding more reasonable results.

\subsubsection{Attentive Knowledge Tracing} \label{attentive}
In the development of deep learning, the Transformer is initially proposed for neural machine translation \citep{vaswani2017attention}, which abandons recurrence and solely relies on the self-attention mechanism to capture global dependencies within a sequence.  The Transformer has been demonstrated to excel in feature extraction and dependency capture, while maintaining high computational efficiency. Some representative pre-training models based on the Transformer, such as BERT \citep{devlin-etal-2019-bert} and GPT \citep{brown2020language}, have obtained state-of-the-art results on various natural language processing tasks. \citet{pandey2019self} propose a self-attentive model for knowledge tracing (SAKT), which directly apply the Transformer to capture long-term dependencies between students' learning interactions. Furthermore, \citet{2021Fine} introduce an adaptive sparse self-attention network to generate missing features and simultaneously produce fine-grained predictions of student performance. \citet{2020Learning} employ a multi-head ProbSparse self-attention mechanism to mitigate the time complexity and effectively capture the long-term dependencies in students' learning interactions.

However, the complexity of the KT task often limits the performance of the aforementioned simple Transformer applications. \citet{SKVMN_plus} introduce a novel approach named Separated Self-Attentive Neural Knowledge Tracing (SAINT) to enhance self-attentive computation for KT adaptability. Specifically, SAINT employs an encoder-decoder structure, with the exercise and answer embeddings being separately encoded and decoded by self-attention layers. The separation of the input allows SAINT to stack self-attention layers multiple times, thus capturing complex relations in student interactions. Subsequently, \citet{SAINT_plus} introduce the SAINT+ model, which integrates two temporal features into SAINT: namely, the time taken to answer each exercise and the interval time between consecutive learning interactions. Both SAINT and SAINT+ have outperformed the SAKT model on the student performance prediction task. 

Additionally, \citet{CAKT} observe that SAKT does not surpass DKT and DKVMN in their experiments. Unlike SAINT and SAINT+, they present a context-aware attentive knowledge tracing (AKT) model. This model integrates the self-attention mechanism with psychometric models, creating a more effective system. AKT is composed of four modules: Rasch model-based embeddings, exercise encoder, knowledge encoder, and knowledge retriever. Specifically, the embedding module employs the classic Rasch model in psychometrics \citep{1980Applications} to construct embeddings for exercises and KCs:
\vspace{-0.2cm}
\begin{equation}
	\begin{aligned}
		\bm{x}_t = \bm{c}_{c_t} + \mu_{e_t} \cdot \bm{d}_{c_t}, \\
	\end{aligned}
	\vspace{-0.2cm}
\end{equation}
where $\bm{c}_{c_t} \in \mathbb{R}^{\bm{D}}$ is the embedding of the KC of this exercise, $\bm{d}_{c_t} \in \mathbb{R}^{\bm{D}}$ is a vector that summarizes the variation in exercises with the related KC, and  $\mu_{e_t} \in \mathbb{R}^{\bm{D}}$ is a scalar difficulty parameter that controls the extent to which this exercise deviates from the related KC.
The exercise-answer tuple $(e_t, a_t)$ is similarly extended using the scalar difficulty parameter for each pair:
\vspace{-0.2cm}
\begin{equation}
	\begin{aligned}
		\bm{y}_t = \bm{q}_{(c_t, a_t)} + \mu_{e_t} \cdot \bm{f}_{(c_t, a_t)}, \\
	\end{aligned}
	\vspace{-0.2cm}
\end{equation}
where $\bm{q}_{(c_t, a_t)} \in \mathbb{R}^{\bm{D}}$ is the KC-answer embedding, $\bm{f}_{(c_t, a_t)} \in \mathbb{R}^{\bm{D}}$ is the variation vector. Through the above embedding, exercises labeled as the same KCs are determined to be closely related while retaining important individual characteristics.
Then, in the exercise encoder, the input is the exercise embeddings $ \{\bm{e}_1, . . . , \bm{e}_t \} $ and the output is a sequence of context-aware exercise embeddings $\{\widetilde{\bm{e}}_1, . . ., \widetilde{\bm{e}}_t\}$. AKT designs a monotonic attention mechanism to accomplish the above process, where the context-aware embedding of each exercise depends on both itself and the previous exercises, i.e., $\widetilde{\bm{e}}_t = f_{enc_1}(\bm{e}_1, . . . , \bm{e}_t)$. Similarly, the knowledge encoder takes exercise-answer embeddings $ \{\bm{y}_1, . . . , \bm{y}_t \} $  as input and
outputs a sequence of context-aware embeddings of the knowledge acquisitions $\{\widetilde{\bm{y}}_1, . . ., \widetilde{\bm{y}}_t\}$ using the same monotonic attention mechanism; these are also determined by students' answers to both the current exercise and prior exercises, i.e., $\widetilde{\bm{y}}_t = f_{enc_1}(\bm{y}_1, . . . , \bm{y}_t)$.
Finally, the knowledge retriever takes the context-aware exercise embedding $\widetilde{\bm{e}}_{1:t}$ and exercise-answer pair embeddings $\widetilde{\bm{y}}_{1:t}$ as input and outputs a retrieved knowledge state $\bm{h}_t$ for the current exercise. Since the student’s current knowledge state depends on answering the related exercise, it is also context-aware in AKT.
The novel monotonic attention mechanism proposed in AKT is based on the assumption that the learning process is temporal and students' knowledge will decay over time. Therefore, the scaled inner-product attention mechanism utilized in the original Transformer is not suitable for the KT task. AKT uses exponential decay and a context-aware relative distance measure to compute the attention weights. Finally, AKT achieves outstanding performance in predicting students' future answers, as well as demonstrating interpretability due to the combination of the psychometric model.

It is important to note that \citet{pu2022self} have recently proposed that attentive knowledge tracing models significantly benefit from students' continuous, repeated interactions on the same exercises throughout the learning process. In their experiments, the removal of these repeated interactions in the dataset led to a decline in AKT's performance, bringing it close to that of DKVMN. 
{Moreover, according to the findings of \citet{yin2023tracing},} existing attentive attentive KT models primarily trace patterns of a learner's learning activities, rather than their evolving knowledge states. Consequently, they develop the DTransformer model to facilitate stable knowledge state estimation and tracing, rather than solely focusing on next performance prediction. 

\begin{figure}[t]

	\centerline{\includegraphics[width=0.9\columnwidth]{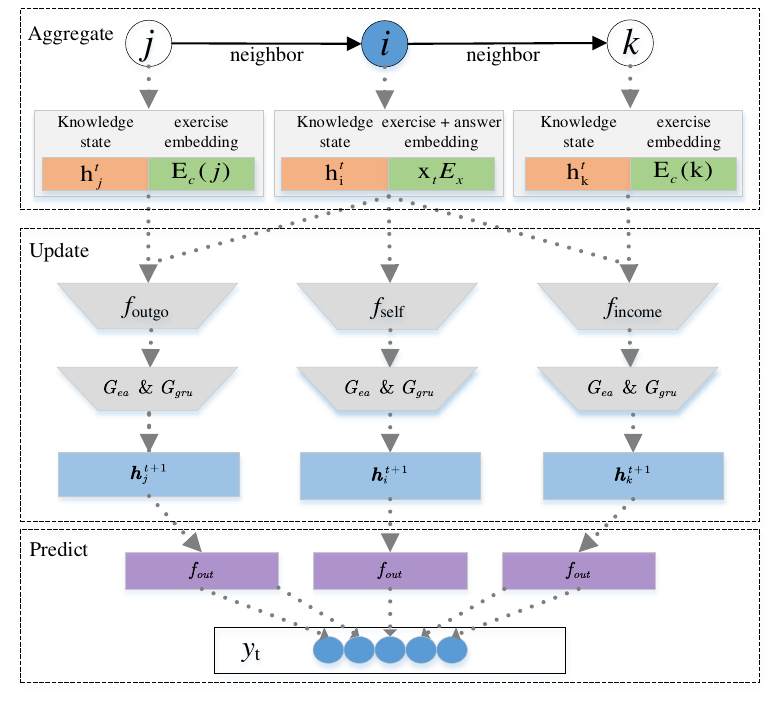}}
	\vspace{-0.3cm}
	\caption{The architecture of graph-based knowledge tracing \citep{nakagawa2019graph}.}
	\label{fgkt}
	\vspace{-0.3cm}
\end{figure}

\subsubsection{Graph-based Knowledge Tracing} \label{graph}
Graph neural networks (GNNs), which are designed to handle complex graph-related data, have developed rapidly in recent years \citep{wu2019comprehensive}. The graph represents a kind of data structure that models a set of objects (nodes) and their relationships (edges). From a data structure perspective, there is a naturally existing graph structure within the KCs. Therefore, incorporating the graph structure of the KCs as additional information should be beneficial to the KT task.
\citet{nakagawa2019graph} presented graph-based knowledge tracing (GKT), which conceptualizes the potential graph structure of the KCs as a graph $G = (V, E)$, where nodes $V = \{v_1, v_2, ..., v_N\}$ represent the set of KCs and the edges $E \subseteq V \times V $ represent relationships of these KCs; moreover, $ \bm{h}^t = \{\bm{h}^t_{i \in V}\} $ represents the student's temporal knowledge state after answering the exercise at time $t$.
The architecture for graph-based knowledge tracing is presented in Figure \ref{fgkt}, which is composed of three parts: (1) $aggregate$, (2) $update$ and (3) $predict$.

In the $aggregate$ module, GKT aggregates the  temporal knowledge state and the embedding for the answered KC $i$ and its neighboring KC $j$:
\vspace{-0.1cm}
\begin{equation}
	\bm{h}_k^{'t}=\left\{
	\begin{aligned}
		&[\bm{h}_k^{t}, a^t\bm{E}_s]  &(k = i), \\
		&[\bm{h}_k^{t}, \bm{E}_e(k)]  &(k \neq i),
	\end{aligned}
	\right.
	\vspace{-0.1cm}
\end{equation}
where $a^t$ represents the exercises answered correctly or incorrectly at time step $t$, $\bm{E}_s$ is the embedding matrix for the learning interactions, $\bm{E}_e$ is the embedding matrix for the KC, and $k$ represents the $k$-th row of $\bm{E}_e$.

In the $update$ module, GKT updates the temporal knowledge state based on the aggregated features and the knowledge graph structure, as follows:
\vspace{-0.1cm}
\begin{equation}
	\begin{aligned}
		&\bm{m}_k^{t+1}=
		\begin{cases}
			&f_{self}(\bm{h}_k^{'t})  (k = i), \\
			&f_{neighbor}(\bm{h}_i^{'t}, \bm{h}_k^{'t})  (k \neq i),\\
		\end{cases}\\
		&\widetilde{\bm{m}}_k^{t+1} = G_{ea}(\bm{m}_k^{t+1}), \\
		&\bm{h}_{k}^{t+1} = G_{gru}(\widetilde{\bm{m}}_k^{t+1}, \bm{h}_{k}^{t}), \\
	\end{aligned}
	\vspace{-0.1cm}
\end{equation}
where $f_{self}$ is the multilayer perceptron, $G_{ea}$ is the same $erase$-followed-by-$add$ mechanism used in DKVMN, and $G_{gru}$ is the gated recurrent unit (GRU) gate \citep{cho2014learning}. Moreover, $f_{neighbor}$ defines the information propagation to neighboring nodes based on the knowledge graph structure.

In the $predict$ module, GKT predicts the student's performance at the next time step according to the updated temporal knowledge state:
\vspace{-0.1cm}
\begin{equation}
	y_k^t = \sigma(\bm{W}_k\bm{h}_k^{t+1} + \bm{b}_k),
	\vspace{-0.1cm}
\end{equation}
where $\bm{W}_k$ is the weight parameter and $\bm{b}_k$ is the bias term.

In addition to modeling the graph structure in KCs by graph neural networks, \citet{lu2022cmkt} propose to model the educational relation and topology in the concept map, which will be intended to act as mathematical constraints for the construction of the KT model.
Recently, in the attempt to further explore knowledge structure, \citet{tong2020structure} propose structure-based knowledge tracing (SKT), which aims to capture the multiple relations in knowledge structure to model the influence propagation among concepts. SKT is mainly motivated by an education theory, \emph{transfer of knowledge} \citep{Royer1979Theories}, which claims that students' knowledge states on some relevant KCs will also be changed when they are practicing on a specific KC due to the potential knowledge structure among KCs. Therefore, a student's knowledge state is determined by not only the temporal effect from the exercise sequence, but also the spatial effect from the knowledge structure. To concurrently model the latent spatial effects, SKT presents the synchronization and partial propagation methods to characterize the undirected and directed relations between KCs, respectively. In this way, SKT  measures influence propagation in the knowledge structure with both temporal and spatial relations. To get rid of dependence on knowledge structure, \citet{long2022automatical} propose the Automatical Graph-based Knowledge Tracing (AGKT), which utilizes the automatical
graph to measure students’ knowledge states automatically without annotation manual annotations.

\subsection{Summarization}

{It is crucial to emphasize that, despite deep learning models exhibiting superior performance compared to Bayesian models and logistic models, there remains significant room for improvement in their interpretability and explainability.  
Due to the end-to-end learning strategy, deep learning models are notoriously difficult to interpret. The modeling process itself is also challenging to explain. Specifically, deep learning models are predominantly data-driven and benefit a lot from large-scale student learning data. It is challenging to understand how they calculate a student's knowledge state with no theoretical guidance \citep{zanellati2024hybrid}. 
All we have are the results generated by these models. Therefore, any errors made by these models will lead students to doubt their reliability. The lack of explainability and interpretability has thus limited their further applicability. }

{To make the complex KT models interpretable, especially those deep learning models, researchers have attempted various methods.
\citet{lu2020towards} presented a post-hoc approach to reveal the interpretability of DKT. Specifically, they employed the layer-wise relevance propagation (LRP) technique \citep{montavon2019layer} to interpret DKT by measuring the relevance between DKT's output and input. Preliminary experimental results suggest that this post-hoc approach could be a promising method for explaining DKT. Besides, explainable AI (xAI) is proposed to make the black-box deep learning models more transparent, thereby promoting its applications \citep{saeed2023explainable}. \citet{lu2023interpreting} proposed to use the xAI technique to interpret the complex KT models based on deep learning. The interpreting results of DKT have been demonstrated to aid in enhancing the trust of students and teachers. Their findings suggested that it is promising to utilize xAI techniques to interpret the deep learning KT models, thereby assisting users in accepting and applying the suggestions provided by these models. 
}

\section{Variants of Knowledge Tracing Models} \label{sec:variants}
{The aforementioned fundamental KT models are} typically based on simplistic assumptions regarding the learning process. Specifically, these models predominantly employ learning interactions such as exercises and responses to estimate students' knowledge states. However, the learning process is not solely represented by exercises and responses, but is influenced by various factors. In summary, the aforementioned fundamental KT models, while straightforward, may have reduced performance in real-world learning scenarios. Consequently, numerous variants have been proposed under more stringent assumptions, reflecting a more comprehensive learning process in real-world scenarios. Accordingly, we classify and review current variants of fundamental KT models into four categories: (1) Modeling individualization before learning, (2) Incorporating engagement \ during learning, (3) Considering forgetting after learning, and (4) utilizing side information across learning.

\subsection{Modeling Individualization before Learning}
Everything and everyone possess unique characteristics. For instance, \citet{liu2011personalized} explored several personalized factors of various tourists to recommend personalized travel packages. Similarly, the concept of individualization in the KT task implies that different students often exhibit different learning characteristics (such as varying learning rates or prior knowledge). Considering the student-specific variability in learning could potentially enhance the KT process, as suggested by \citet{yudelson2013individualized}. In the subsequent sections, we will introduce various variant KT models that {take into account individualization before learning.}

The initial BKT paper has delved into the concept of individualization. Specifically, it uses all students' learning interactions on a specific KC to learn the individual parameter. Similarly, for a specific student, all her learning interactions are utilized to fit her individual learning parameters \citep{corbett1994knowledge}. Consequently, BKT is able to ascertain different learning and performance parameters for various students and KCs. However, this approach only offers a marginal improvement compared to original BKT.  

Subsequently, \citet{pardos2010modeling} propose two simple variants of BKT that respectively individualize students' initial probability of mastery and the probability of transition from the unlearned state to the learned state. {Specifically, a student node is added to individualize the initial probability of mastery for each student. The student node assigns each student with a personalized initial probability of mastery. A conditional probability table is designed to determine the value of the student node. Similarly, if changing the connection of the student node to the subsequent knowledge nodes, the transition probability parameter can also be individualized. In this case,  the student node gives individualized transition parameters to each student.}
Moreover, rather than individualizing only one kind of parameter in BKT, some other variants of BKT opt to individualize all four BKT parameters simultaneously \citep{yudelson2013individualized}. \citet{lee2012impact} suggest that when applied in an intelligent tutoring system, the individualized BKT model can yield good improvements to student learning efficacy, reducing by about half the amount of questions required for 20\% of students to achieve mastery.

Another means of modeling individualization is clustering, which considers a wider range of students in different groups \citep{pardos2012clustered}. 
By clustering the students into various groups, we can train different KT models and make predictions on the test data. The number of clusters is then varied according to the student groups and the predicting process is repeated iteratively. Finally, we can obtain a set of different predictions. Furthermore, there are two common methods used to combine these predictions \citep{trivedi2011clustering}: (1) uniform averaging, which simply averages the predictions; (2) weighted averaging, which combines the models by means of a weighted average.
To realize clustering, K-means is a basic clustering algorithm that randomly initializes a set of cluster centroids, which are identified using Euclidean distance. Another popular clustering algorithm is spectral clustering, which represents the data as an undirected graph and analyzes the spectrum of the graph Laplacian obtained from the pairwise similarities of data points. Recently, some novel clustering algorithms have been proposed, including discrete nonnegative spectral clustering \citep{7920371} and clustering uncertain data \citep{6051435}.

\citet{minn2018deep} propose a model named deep knowledge tracing with dynamic student classification (DKT-DSC), which introduces individualization to DKT by exploiting the idea of clustering. 
According to students' previous performance, DKT-DSC assigns students with similar learning ability to the same group . The knowledge states of students in different groups are then traced by different DKT models. Moreover, considering the dynamic property of the learning ability, each student's learning sequence is segmented into multiple time intervals. At the start of each time interval, DKT-DSC will reassess students' learning ability and reassign their groups. 
In DKT-DSC, the K-means clustering algorithm is utilized to split students with similar ability levels into the same group at each time interval. After learning the centroids of all K clusters, each student is assigned to the nearest cluster. Through dynamic student clustering, DKT-DSC offers an effective approach to realizing individualization in DKT.

\citet{10076899} claim that it is significant to consider both individual exercise representation and individual prior knowledge. They propose a fine-grained knowledge tracing model, named FGKT. FGKT obtains the individual exercise representation through the acquisition of knowledge cells (KCs) and exercise distinctions. Subsequently, it assesses the individual prior knowledge by evaluating the relevance between current and historical learning interactions. Finally, the above individual representations will be utilized as the input of LSTM in FGKT to evaluate students' evolving knowledge states. \citet{zhao2023exploiting} also notice that the individualization of exercises is significant for measuring students' knowledge states. They propose to consider multiple exercise factors, including the difficulty
and the discrimination, to enhance the performance of DKT.

\citet{shen2020convolutional} propose a convolutional knowledge tracing model (CKT) to implicitly measure student individualization. Specifically, CKT considers two factors that influence students' individualization: individualized learning rates and individualized prior knowledge.
Individualized learning rates represent students' differing capacities to absorb knowledge. The sequence of student learning interactions can reflect different learning rates in the sense that students with high learning rates can rapidly master knowledge, while others need to spend more time trying and failing. Therefore, it is reasonable to assess the differences in learning rate by simultaneously processing several continuous learning interactions within a sliding window of convolutional neural networks \citep{lecun2015deep}. Besides, individualized prior knowledge refers to students' prior knowledge, which can be assessed via their historical learning interactions.

\subsection{Incorporating Engagement during Learning}

Student engagement is defined as \emph{"the quality of effort students themselves devote to educationally purposeful activities that contribute directly to desired outcomes"} \citep{trowler2010student}. This definition highlights a strong connection {between student engagement and the learning process.} Generally, higher engagement leads to enhanced knowledge gains. Consequently, considering student engagement in the learning process could potentially improve knowledge tracing results \citep{carini2006student}. In this section, {we will present some variants that integrate student engagement into the KT models. }

Student engagement is difficult to be directly measured. In practice, some online learning systems have made use of sensor data to measure student engagement. For example, inexpensive portable electroencephalography (EEG) devices can help to detect a variety of student mental states in learning, which can be seen as reflections of student engagement \citep{berka2007eeg}. \citet{xu2014eeg} propose two methods that combine EEG-measured mental states to improve the performance of BKT. Concretely, the first one inserts a one-dimensional binary EEG measure into BKT, forming the EEG-BKT structure that extends BKT by adding a binary variable node $E$ between the knowledge node and the answer node. The second one, i.e., EEG-LRKT, utilizes logistic regression to combine an $m$-dimensional continuous variable $E$ extracted from the raw EEG signal in BKT.

However, in most cases, it is difficult to collect sensor data on every student. Therefore, \citet{schultz2014tracing} propose the knowledge and affect tracing (KAT) to model both knowledge and engagement in parallel. KAT is a sensorless model that does not rely on any sensor data. In this model, both knowledge and engagement are assumed to have direct influences on student performance. KAT considers three kinds of disengagement behaviors: quick guess (the student makes an attempt very quickly), bottom-out hint (all available hints are used) and many attempts (making more than three attempts at an exercise). These three behaviors are grouped as “gaming” behaviors in order to predict students' knowledge and engagement at each learning interaction. 
Rather than assuming equal influence of knowledge and engagement on students' knowledge state, one variation on the KAT model defines the connection between knowledge and engagement, and accordingly considers that students' knowledge states will influence their engagement. For example, students are more likely to disengage from knowledge they are not familiar with.
Moreover, rather than explicitly modeling student engagement, \citet{schultz2014modeling} further propose the knowledge tracing with behavior (KTB) model, which has only one latent knowledge node that acts as a combination of both knowledge and engagement. KTB assumes that both engagement and performance are expressions of knowledge. The Bayesian estimation of the knowledge state needs to be inferred by both student engagement and performance.

\citet{mongkhonvanit2019deep} propose to add five features in the process of watching videos on MOOCs to the input of DKT. These features reflect student engagement from various aspects, including playback speed, whether or not the video was paused, fast-forwarded or rewound, and whether or not the video was completed. For example, if a student watches a video at a much faster playback speed, it is likely that he/she is impatient and absent-minded.  This model incorporates two further features: whether or not the exercise was submitted with an answer selected and whether or not the exercise was a part of an end-of-unit quiz, both of which are considered together. Experimental results indicate that DKT can achieve better performance through incorporating the above binarized engagement covariates.

\subsection{Considering Forgetting after Learning}
In real-world scenarios, while learning, forgetting is inevitable \citep{markovitch1988role}. The \emph{Ebbinghaus forgetting curve theory} indicates that students' knowledge proficiency will decline due to forgetting \citep{chen2017tracking}. Recently, \citet{huang2020learning} proposed the concept of 'Knowledge Proficiency Tracing' (KPT), a model that can dynamically capture the changes in students' proficiency levels on knowledge concepts over time. This model effectively tracks these changes in an interpretable manner. Therefore, the assumption that students' knowledge states will remain constant over time is untenable. However, fundamental KT models, such as the BKT, often overlook forgetting. In the following, we will introduce some variants of fundamental KT models that have attempted to consider forgetting after learning for more precise knowledge states. 

\citet{qiu2011does} discover that BKT consistently overestimates the accuracy of students' answers when a day or more had elapsed since {their previous responses.} The underlying reason is that BKT assumes that student performance will remain the same regardless of how much time has passed. To consider how student performance declines with time, they propose a BKT-Forget model, which hypothesizes that students may forget information they have learned as days go by. In the BKT-Forget model, a time node is added to specify which parameters should be affected by a new day and the new day node is fixed with a prior probability of 0.2. {It also introduced parameters to represent the forgetting rate on a new day and denote the forgetting rate on the same day.} However, although BKT-forget does consider the decline in student performance, it can only model forgetting that occurs over the time scale of days. 
To model the continuous decay of knowledge as time progresses, \citet{nedungadi2015incorporating} incorporate forgetting into BKT based on the assumption that learned knowledge decays exponentially over time \citep{loftus1985evaluating}. An exponential decay function is thus utilized to update the knowledge mastery level. They further assumed that the chance of forgetting will increase if a student does not practice the knowledge concepts within 30 days.
Moreover, \citet{khajah2016deep} introduce an approach that counts the number of intervening trials and treats each as an independent opportunity for forgetting to occur.

Recall the PFA model in Eq.(\ref{pfa}), in which the probability of students' mastery is estimated using a logistic function: $ p(\theta) = \sigma(\beta + \mu s + \nu f)$. The original PFA model ignores the order of answers, in addition to the time between learning interactions. It is therefore difficult to directly incorporate time information into the original PFA model.
\citet{1199904} propose PFAE (PFA Elo/Extended), a variant of the PFA model that combines PFA with some aspects of the Elo rating system \citep{Pel2014Time}. The Elo rating system is originally devised for chess rating (estimating players' skills based on match results). In PFAE, $\theta$ is updated after each learning interaction:
\vspace{-0.1cm}
\begin{equation}
	\theta :=\left\{
	\begin{aligned}
		& \theta + \mu \cdot (1 - p(\theta))& \mbox{if the answer was correct}, \\
		& \theta + \nu \cdot p(\theta)&  \mbox{if the answer was wrong}.
	\end{aligned}
	\right.
	\vspace{-0.1cm}
\end{equation}

As the forgetting behavior of students is closely related to time, in order to consider forgetting, \citet{pelanek2015modeling} add a time effect function $f$ to $\theta$, i.e., using $p(\theta+f(t))$ instead of $p(\theta)$, where $t$ is the time (in seconds) from the last learning interaction, and $f$ is the time effect function.

To represent the complex forgetting behavior, the DKT-forget model \citep{nagatani2019augmenting} introduces forgetting into DKT, which considers three types of side information related to forgetting: (1) the repeated time gap that represents the interval time between the present interaction and the previous interaction with the same KC, (2) the sequence time gap that represents the interval time between the present interaction and the previous interaction, and (3) past trial counts that represent the number of times a student has attempted on the exercise with the same KC. All these three features are discretized at $log_2$ scale.
Those side information is concatenated as additional information and represented as a multi-hot vector $\bm{c}_t$, which is integrated with the embedding vector $\bm{v}_t$ of the learning interaction, as follows:
\vspace{-0.1cm}
\begin{equation}
	\bm{v}_t^c = \theta^{in}(\bm{v}_t, \bm{c}_t),
	\vspace{-0.1cm}
\end{equation}
where $\theta^{in}$ is the input integration function. The integrated input $\bm{v}_t^c$  and the previous knowledge state $\bm{h}_{t-1}$ are passed through the RNNs to update $\bm{h}_t$ in the same way as in Eq.(\ref{dkt}). The additional information at the next time step $\bm{c}_{t+1}$ is also integrated with the updated $\bm{h}_t$:
\vspace{-0.1cm}
\begin{equation}
	\bm{h}_t^c = \theta^{out}(\bm{h}_t, \bm{c}_{t+1}),
	\vspace{-0.1cm}
\end{equation}
where $\theta^{out}$ is the output integration function.

\citet{hawkesKT} propose a novel HawkesKT model, which introduces the Hawkes process to adaptively model temporal cross-effects. The Hawkes process performs well at modeling sequential events localized in time, as it controls corresponding temporal trends by the intensity function. The intensity function in HawkesKT is designed to characterize the accumulative effects of previous learning interactions, along with their evolutions over time. In HawkesKT, the temporal cross-effects and the ways in which they evolve between historical learning interactions combine to form a dynamic learning process.

\subsection{Utilizing Side Information across Learning} \label{sec:side information}

Most KT models primarily {rely on exercises and student responses} to evaluate students' knowledge states. These models have yielded impressive results and have been effectively implemented in online learning systems. Despite this, there are various other types of side information collected across the learning process that could be utilized to enhance {these models.} In this section, we will introduce several variants that aim to leverage this diverse side information across learning.

In terms of a student's first response time, a short initial response time could indicate either high proficiency or 'gaming' behavior, while a long initial response time could indicate either careful thinking or lack of concentration. Since the connection between initial response time and knowledge state could be influenced by {complex factors}, \citet{wang2012leveraging} propose to discretize the continuous first response time into four categories (i.e., extremely short, short, long, extremely long) {to eliminate unnecessary information and simplify the latent complex possibilities. They then build a one-by-four parameter table, in which each column represents the category of the initial response time of the previous exercise, while the relevant values represent the probability of correct answers}.

Regarding tutor intervention, \citet{beck2008does} propose the Bayesian evaluation and assessment model, which simultaneously assesses students' knowledge states and evaluates the lasting impact of tutor intervention. More specifically, it adds one observable binary intervention node to BKT: \emph{True} means that the tutor intervention occurs in corresponding interactions while \emph{False} indicates the opposite. The connection between the intervention node and knowledge node indicates the potential impact of the tutor intervention on students' knowledge states. The intervention node is linked to all four BKT parameters. As a result, there are a total of eight parameters to learn in order to incorporate tutor intervention. One possible way to reduce the number of parameters is choosing to link only the intervention node to the learning rate parameter \citep{sao2013incorporating}.
Similarly, \citet{lin2016intervention} develop the intervention-Bayesian knowledge tracing (Intervention-BKT) model, {which incorporates two types of interventions into BKT and distinguishes their different effects: \emph{elicit and tell}}. The relations between the intervention and performance nodes represent the impact of teaching interventions on student performance, while the relations between the intervention and knowledge nodes represent the impact of teaching interventions on students' knowledge states. Therefore, at each learning interaction, while the present knowledge state is conditional on both the previous knowledge state and the current intervention, the student's performance depends on both the present knowledge state and the current intervention.

Rather than considering only one kind of side information, \citet{gonzalez2014general} propose a feature-aware student knowledge tracing (FAST) model, which allows for the utilization of all kinds of side information. Traditional BKT uses conditional probability tables for the guessing, slipping, transition and learning probabilities, meaning that the number of features involved in inference grows exponentially. Therefore, as the number of features increases, the time and space complexity of the model also grow exponentially. To deal with this large number of features, FAST uses logistic regression parameters rather than conditional probability tables. The number of features and complexity increase linearly rather than exponentially.  
For parameter learning, FAST uses the Expectation Maximization with Features algorithm \citep{berg2010painless} and focuses on only emission features. The E step uses the current parameter estimates $\lambda$ to infer the probability of the student having mastered the KC at each learning interaction.
The parameters  $\lambda$ are now a function of the weight $\beta$ and the feature vector $\bm{f}(t)$.  $\bm{f}$ is the feature extraction function, and  $\bm{f}(t)$ is the feature vector constructed from the observations at the relevant time step. The emission probability is represented with a logistic function:
\vspace{-0.1cm}
\begin{equation}
	\begin{aligned}
		\lambda(\beta)^{y^{'},k^{'}} & = \frac{1}{1 + exp(-\beta^T \cdot \bm{f}(t))},
	\end{aligned}
	\vspace{-0.1cm}
\end{equation}
where $\beta$ is learned by training a weighted regularized logistic regression using a gradient-based search algorithm.

\citet{zhang2017incorporating} propose an extension to DKT that explored the inclusion of additional features. Specifically, it incorporates an auto-encoder network layer to convert the higher-dimensional input data into smaller representative feature vectors, thereby reducing both the resource and time requirement for training. Students' response time, opportunity count, and first action are selected as incorporated side information and all input features are converted into a fixed-length input vector. First, all input features are converted into categorical data and represented as a sparse vector by means of one-hot encodings. These encoded features are concatenated together to construct the higher-dimensional input vector:
\vspace{-0.2cm}
\begin{equation}
	\begin{aligned}
		C(e_t,a_t)  &= e_t + (max(e) +1)a_t, \\
		v_t &=  O(C(e_t,a_t)) \oplus O(C(t_t,a_t)) \oplus O(t_t), \\
		v_t^{'} &=  tanh(W_vv_t + b_v),
	\end{aligned}
	\vspace{-0.2cm}
\end{equation}
where $C$ is the cross feature, $O$ is the one-hot encoder format, $v_t$ represents the resulting input vector of each learning interaction, $e_t$ is the exercise, $a_t$ refers to the answer, $t_t$ is the response time, $W_v$ is the weight parameter and $b_v$ is the bias term. Subsequently, an auto-encoder is introduced to reduce the dimensionality without incurring the loss of too much important information. Finally, the feature vectors extracted by auto-encoder will be the new input of DKT.

\citet{huang2019ekt} presented another extension to DKT, namely the Exercise-aware Knowledge Tracing (EKT), which utilized the potential value of exercises' text contents. Generally, the text content is of great significance for students to understand the exercises. For example, \citet{huang2017question} used text materials to automatically predict their difficulties, \citet{liu2018finding} utilized the text content to find similar exercises. \citet{yin2019quesnet} further proposed a pre-training model called QuesNet for learning the unified representations of heterogeneous exercises. 
Therefore, instead of using one-hot encoding of exercises, EKT automatically learns the semantic representation of each exercise from its text contents. EKT first uses $Word2vec$ \citep{mikolov2013distributed} to pre-train the embedding vector for each word in the exercise. It then constructs a bidirectional LSTM, which captures the word sequence from both forward and backward directions to learn the semantic word representation. The element-wise max-pooling operation is utilized to merge words’ contextual representations into a global embedding. Finally, EKT can update the student's knowledge state with the aid of the semantic representation of each exercise.

To achieve more feasible integration of side information, \citet{loh2011classification} present a deep knowledge tracing method with decision trees (DKT-DT), which takes advantage of Classification And Regression Trees (CART) to preprocess the heterogeneous input features \citep{cheung2017heterogeneous}. Specifically, CART is utilized to automatically partition the feature space and outputs whether or not a student can answer an exercise correctly. {The predicted response and the true response are encoded into a four-bit binary code; for example, the code is $1010$ if the predicted response and the true response are both correct. This binary code is then concatenated with the original one-hot encoding of the exercise as the new input of DKT to train the corresponding model.}

\citet{jung2023language} suggests that the student's language proficiency can serve as supplementary information to improve existing KT models. The student’s language proficiency is extracted by Elo rating score and time window features. Then, the language proficiency information is demonstrated to be effective in promoting several KT models, including DKT, DKVMN, and SAKT. Additionally, the problem of cold start in the KT task is alleviated with the assistance of language proficiency information. 
\citet{liu2023enhancing} explore to add side information to the original KT model by auxiliary learning tasks. They specifically introduced two tasks: (1) predicting the KCs of the question, and (2) predicting the individualized prior knowledge. By training with these tasks, KT can enhance its understanding of the intrinsic relationships between questions and KCs, while explicitly capturing student-level variability.

{When solving programming problems, we can record students' full code submissions, which can be employed to analyze their programming ability. \citet{kasurinen2009estimating} collected student programming data and analyzed their personalized programming preferences. The study commenced with a statistical analysis of student errors, followed by an examination of students' programming structures derived from their code submissions, and finally used BKT to measure students' programming abilities.
\citet{wang2017deep} transformed students' code submissions into embedded vectors, and applied them in DKT to model students' fine-graded programming knowledge states.
\citet{zhu2022enhancing} noticed that a single programming problem generally involves in multiple KCs, thereby they proposed to learn useful information about the programming problem's multiple requirements from students' code submissions. }

\section{Applications} \label{sec:applications}
Although knowledge tracing is an emerging research area, it has already been applied in a wide variety of scenarios. {In the following, we will first survey the applications of KT models in two typical educational scenarios: learning resources recommendation and adaptive learning. Then, we will discuss broader applications of KT beyond student learning.}

\subsection{Learning Resources Recommendation}

Traditionally, learning resources for each student are selected in one of two ways. The first one requires teachers to manually select suitable resources that match students' knowledge levels. However, this approach requires substantial time and effort, and different teachers may have different preferences. The second one allows students themselves to freely choose resources to learn. However, this may result in students choosing too easy or too difficult materials that will not benefit their learning \citep{Desmarais2012}, leading to low learning efficiency. In recent years, the prevalence of intelligent tutoring systems and the development of KT methods have made it possible to automatically recommend appropriate exercises to each student based on artificially designed algorithms.

Exercises are the most common learning resources in learning. Given the inferred knowledge states, one common strategy is selecting the next exercise that will best advance students' knowledge acquisition. \citet{Desmarais2012} propose two extensions of the original BKT model, which respectively considered  exercises' difficulties and students' multiple-attempt behaviors. These two extensions are integrated into a BKT-sequence algorithm to recommend exercises to students based on their knowledge states. Specifically, BKT-sequence first determines the predicted range of scores for each exercise. It then computes an expected score for each exercise that the student should get to achieve mastery, which is dependent on their current knowledge state (for instance, a lower knowledge state will result in higher expected scores). Finally, the algorithm returns the exercise with a predicted score that is closest to that of the expected score. Therefore, as the knowledge state of a particular KC grows, more difficult exercises will be recommended, as harder exercises are associated with a lower predictive score. Experimental results have shown that students using the BKT-sequence algorithm were able to solve more difficult exercises, obtained higher performance and spent more time in the system than students who used the traditional approach. Moreover, students also expressed that the BKT-sequence algorithm was more efficient. \citet{10038578} expanded the DKVMN model \citep{zhang2017dynamic} to include the exercise's type and difficulty. This model is then used to assess students' knowledge state and subsequently recommends personalized exercises for each student in Ssmall Private Online Courses (SPOCs). They conducted a randomized controlled trial to show the proposed personalized exercise recommendation could enhance students' learning efficiency.

In addition to exercises, there are also some other types of multi-modal learning resources, such as videos and figures. \citet{Machardy2015} utilizes an adaptation of BKT to improve student performance prediction by incorporating video observation. Experimental verification demonstrates the impact of both using and eschewing video data, as well as the learning rate associated with a particular video. In this way, they further developed a method to help people evaluate the quality of video resources. Concretely, they proposed the Template 1 Video model to incorporate video observations into BKT, which adds video activity as additional independent observation nodes to the BKT model. This model accordingly considers the probability that a given video resource will impart knowledge to a student. Moreover, the transition probability in BKT is conditional only on the presence of either a video or an exercise. Thus, the quality of the video can be determined by its promotion of learning, and this model can be leveraged as a tool to aid in evaluating and recommending video resources.

When recommending learning resources, the primary aim of existing solutions is to choose a simple strategy for assigning non-mastered exercises to students. 
While reasonable, it is also too broad to advance learning effectively. \citet{Zhenya2019Exploring} accordingly propose three more beneficial and specific objectives: \emph{review and explore}, \emph{smoothness of difficulty level} and \emph{student engagement}. In more detail, \emph{review and explore} considers both enhancing students' non-mastered concepts with timely reviews and reserving certain opportunities to explore new knowledge; \emph{smoothness of difficulty level} indicates that the difficulty levels of several continuous exercises should vary within a small range as students gradually learn new knowledge; finally, \emph{student engagement} considers that to promote students' enthusiasm during learning, the recommended exercises should be in line with their preferences. In order to support online intelligent education with the above three domain-specific objectives,
they developed a more reasonable multi-objective deep reinforcement learning (DRE) framework. DRE presented three corresponding novel reward functions to capture and quantify the effects of the above three objectives. This DRE framework is a unified platform designed to optimize multiple learning objectives, where more reasonable objectives also can be incorporated if necessary.
Experimental results show that DRE can effectively learn from the students' learning records to optimize multiple objectives and adaptively recommend suitable exercises.

\subsection{Adaptive Learning}

{Adaptive learning, unlike learning resource recommendations, goes beyond the mere provision of resources. It not only concentrates on the selection of appropriate learning materials but also designs effective learning strategies and dynamic learning pathways. These are structured based on both the learning rules and students' evolving knowledge states.
Specifically, adaptive learning broadly refers to \emph{"a learning process in which the content taught, or the way such content is presented, changes or 'adapts' based on individual student responses, and which dynamically adjusts the level or types of instruction based on individual student abilities or preferences"} \citep{Oxman2014}. }

The first few attempt made to apply KT to adaptive learning was the ACT Programming Tutor (APT) \citep{corbett1994knowledge}, where students were asked to write short programs and BKT was utilized to estimate their evolving knowledge state. This tutor can present an individualized sequence of exercises to each student based on their estimated knowledge states until the student has "mastered" each rule. 

In recent years, Massive Open Online Courses (MOOCs) have become an emerging modality of learning, particularly in higher education. \citet{pardos2013adapting} adapt BKT on the edX platform. The research object was a 14-week online course that included weekly video lectures and corresponding lecture problems. BKT was applied to enhance students' learning on this course. In order to better adapt BKT to the learning platform, the original BKT was modified in several respects. First, due to the lack of labeled KCs, the problems would be directly seen as the KCs, while the questions would be seen as the exercises belonging to the KC.
Second, in order to capture the varying degrees of students' knowledge acquisition at each attempt, the modified model assigned different guess and slip parameters to different attempt counts. Third, to deal with the problem of multiple pathways in the system, which reflected that the impacts on learning may come from various resources, they framed the influence of resources on learning as a credit/blame inference problem.

Generally, students' cognitive structures include both students' knowledge level and the knowledge structure of learning items (e.g., \emph{one-digit addition} is the prerequisite knowledge of \emph{two-digit addition}). Therefore, adaptive learning should maintain consistency with both students' knowledge level and the latent knowledge structure. Nevertheless, existing methods for adaptive learning often focus separately on either the knowledge levels of students (i.e., with the help of specific KT models) or the knowledge structure of learning items. To fully exploit the cognitive structure for adaptive learning, \citet{liu2019exploiting} propose a Cognitive Structure Enhanced framework for adaptive Learning (CSEAL). CSEAL conceptualized adaptive learning as a Markov Decision Process. It first utilized DKT to trace the evolving knowledge states of students at each learning step. Subsequently, the authors designed a navigation algorithm based on the knowledge structure to ensure that the learning paths in adaptive learning were logical and reasonable, which also reduced the search space in the decision process. Finally, CSEAL utilized the actor-critic algorithm to dynamically determine what should be learned next. In this way, CSEAL can sequentially identify the most suitable learning resources for different students.

\subsection{Broader Applications}
{The above two types of applications are most commonly used for KT in student learning. In addition, the KT methods can be expanded to be utilized in any systems that necessitate continuous evaluation of user capabilities or states. We will introduce some broader applications of KT in this section.}

In gaming systems, the paradigm of tracing students' knowledge state can also work for player modeling. Here, player modeling, which is the study of computational models of players in games, aims to capture human players' characteristics and cognitive features \citep{Yannakakis2013Player}. For instance, \citet{fisch2011children} reveal that children engage in cycles of increasingly sophisticated mathematical thinking over the course of playing an online game. \citet{kantharaju2018tracing} present an approach to trace player knowledge in a parallel programming educational game, which is capable of measuring the current players' real-time state across the different skills required to play an educational game based only on in-game player activities. \citet{2017Educational} conduct a classroom experiment comparing a commercial game for equation solving, i.e., \emph{DragonBox}, with a research-based intelligent tutoring system, i.e., \emph{Lynnette}. The results indicated that students who used \emph{DragonBox} enjoyed the experience more, while students who used \emph{Lynnette} performed significantly better on the test. Therefore, it is possible to enable students to learn effectively and happily by designing suitable educational games on the online learning platform.

{In crowdsourcing, unlabeled data or specific tasks are assigned to various crowd annotators. Understanding the dynamic capabilities of these annotators is crucial in ensuring the reliability of their annotations and promoting the annotation efficiency. \citet{wang2020predicting} developed a framework called KT4Crowd, which utilized KT methods to predict the performance of annotators, which surpassed traditional rating systems. 
Besides, \citet{abdi2020modelling} observed that students' participation in crowdsourcing tasks can enhance their learning. KT methods can also better comprehend  students' knowledge states, aided by the students' annotated items. 
}

\citet{8812979} develop an online citizen science project that employs machine learning techniques to improve the training of new volunteers using authentic tasks featuring uncertain outcomes, such as image classification. Specifically, they employ the BKT model to monitor the knowledge states of volunteers, enabling them to more efficiently complete assigned tasks and contribute meaningfully to the project. 

\citet{zhao2022exercise} develop an automated exercise collection approach for teachers, employing the KT model and reinforcement learning. Specifically, the exercise collection need to be well-designed to align with students' abilities. This study first leverages the KT model to forecast students' performance on unseen exercise candidates.  Subsequently, the exercise selector is designed based on the KT model's predictions, ensuring that the exercise collection is both approximate and optimized. Similarly, \citet{shang2023reinforcement} design a reinforcement learning guided
method for exam paper generation, where the DKT model is utilized to measure examinees' knowledge states.

\section{Datasets and Baselines} \label{sec:dataset}

After introducing above KT models and variants, to better help researchers and practitioners who want to further conduct related work and promote the application of KT, we have open sourced two algorithm libraries, i.e., EduData that for downloading and preprocessing most existing KT-related datasets, and EduKTM that includes extensible and unified implementations of existing popular KT models. In the following, we will give detailed introduction of these two algorithm libraries.

\begin{table*}[t]
	\vspace{-0.6cm}
	\renewcommand\arraystretch{1.2}
	\caption{{Basic information and statistics of existing datasets available for evaluating KT models. }}
	\vspace{-0.2cm}
	\centering
	\resizebox{\textwidth}{!}
	{
		\begin{tabular}{l|c|c|c|c|c|r|r|r|r}
			\hline
			\multirow{3}*{Datasets}&\multirow{3}{*}{Subjects} &\multirow{3}{*}{Learning Stages} & \multirow{3}*{\makecell[c]{Side \\ Information}}& \multicolumn{2}{c|}{Sources}&\multicolumn{4}{c}{Statistics} \\
			\cline{5-10}  
			& & & & \makecell[c]{Online \\ platforms} &  \makecell[c]{Educational \\ challenges} & \makecell[c]{\# of \\ Students} & \makecell[c]{\# of \\ Exercises} & \makecell[c]{\# of \\ KCs} & \makecell[c]{\# of  \\ Learning records}\\
			\hline
			ASSISTments2009 \citep{feng2009addressing}& mathematics & middle school  & Yes & \Checkmark &  & 4,163    & 17,751& 123 & 346,860   \\
			\hline
			ASSISTments2012 \citep{2013Affective}& mathematics & middle school & Yes & \Checkmark &  & 46,674    & 179,999& 265 &6,123,270 \\
			\hline
			ASSISTments2015 & mathematics & middle school & No & \Checkmark &  & 19,917   & /& 100  & 708,631  \\
			\hline
			
			ASSISTments2017& mathematics &\makecell[c]{from middle school \\ to college}  & Yes &  & \Checkmark & 1,709  &3,162& 102 & 942,816 \\
			\hline
			Junyi \citep{ChangHC15} & mathematics & \makecell[c]{from primary \\ to high school} & Yes & \Checkmark &  & 247,606   & 722 & 41 & 25,925,922 \\
			\hline
			Eedi2020 \citep{wang2020diagnostic}& mathematics & \makecell[c]{from primary \\ to high school} & Yes &  & \Checkmark & 118,971    & 27,613 & 388 & 15,867,850  \\
			\hline
			Statics2011 \citep{2010statistics}&engineering &university &Yes & \Checkmark &  & 335    & 1,224 & 80 & 361,092 \\
			\hline
			
			EdNet-KT1 \citep{ednet} & english & / & Yes & \Checkmark &  &  784,309  & 13,169 & 188 &  95,293,926 \\
			\hline
			EdNet-KT2 \citep{ednet} & english &/ & Yes & \Checkmark &  &  297,444  & 13,169 & 188 &  56,360,602 \\
			\hline
			EdNet-KT3 \citep{ednet} & english &/ & Yes & \Checkmark &  &  297,915  & 13,169 & 293 &  89,270,654 \\
			\hline
			EdNet-KT4 \citep{ednet} & english &/& Yes& \Checkmark &  &  297,915  & 13,169 & 293 &  131,441,538 \\
			\hline
			CodeWorkout \citep{edwards2017codeworkout} & programming & university  & Yes &  & \Checkmark & 819    & 50 & 50 &  \textgreater 130,000 \\
			\hline
			
		\end{tabular}
	}
	
	\label{data}
	\vspace{-0.6cm}
\end{table*}

\subsection{Datasets}
As we have mentioned, KT emerges from the development of online education, where a large number of students' learning data is collected for analyzing their learning behaviors and knowledge states. In this section, we mainly introduce existing public datasets available for evaluating KT models. Table \ref{data} lists all datasets, as well as their basic information and statistics. In our released EduData, we provide the service of downloading and preprocessing all these datasets, which is convenient to help beginners analyze and utilize them quickly.
In summary, these datasets are collected in different learning scenarios, so that they exhibit an extremely distinct difference in data scale, subject, and so on, indicating complex applications in practical for KT models. 

\subsubsection{ASSISTments Datasets}
ASSISTments \citep{feng2009addressing}, created in 2004, is an online tutoring system in the United States, which provides students with both assessment information and tutoring assistance. While working on ASSISTments, students will be provided with instructional assistance to help them solve the problem in several substeps when they give wrong answers. After obtaining correct answers, they will be given a new one. Meanwhile, the system will learn about the students' knowledge states and predict how they will do in future tests. Up to now, the organizers have released four public avaliable datasets from ASSISTments, which are respectively ASSISTments2009, ASSISTments2012\footnote{https://sites.google.com/site/assistmentsdata/datasets/2012-13-school-data-with-affect}, ASSISTments2015\footnote{https://sites.google.com/site/assistmentsdata/datasets/2015-assistments-skill-builder-data}, and ASSISTments2017\footnote{https://sites.google.com/view/assistmentsdatamining/dataset}. The ASSISTments datasets have built profound impacts in the research community of KT, which can be seen in many related papers.
Most of these datasets were collected from mathematics in middle school, the details of them are listed as follows.
\begin{itemize}[leftmargin=*]
	\item{\textbf{ASSISTments2009.}}
	The full name of ASSISTments2009 is the ASSISTments 2009-2010 skill builder data set \citep{feng2009addressing}, which was collected form ASSISTments’ skill builder problem sets during the school year from 2009 to 2010. Students were asked to work on exercises with similar knowledge concepts until they can answer correctly for three or more times in a row. This dataset contains many valuable side information, such as \textit{attempt count} that represents the number of attempts of the student, \textit{ms first reponse} that represents the time in the milliseconds for the student’s first response, and \textit{opportunity} that represents the number of opportunities the student has to practice. It worth noting that the original version of ASSISTments2009 has three serious problems that lead to unreliable experimental results \citep{xiong2016going}: (1) a large number of duplicated records, (2) treating scaffolding problems as the same as main problems, and (3) repeated response sequences with different KCs. The latest version of ASSISTments2009 has fixed these problems.
	
	\item{\textbf{ASSISTments2012.}}
	ASSISTments2012 is collected from the ASSISTments system, in the school year from 2012 to 2013 with affect predictions. Compared with ASSISTments2009, ASSISTments2012 has much more students, exercises, as well as learning records. However, many learning records have missed the related KCs in this dataset. After filtering these records, there are only 29,018 students, 53,091 exercises, and 2,711,813 learning records.
	Besides, this dataset contains more side information, such as \textit{start time}, \textit{end time}, \textit{problem type}, and \textit{average confidence}. 
	It worth noting that you can access to the text of the exercises in ASSISTments2012 for more fine-grained research. Specifically, you can email nth@wpi.edu and cc td@wpi.edu, explaining your purpose and promising not to share it with anyone else.
	
	\item{\textbf{ASSISTments2015.}}
	ASSISTments2015 covers the 2015 school years' student response records on ASSISTments. This dataset only contains student learning records on 100 KCs, there is no information about the exercise, as well as any other side information. 
	
	\item{\textbf{ASSISTments2017.}}
	ASSISTments2017 is the dataset used in the 2017 ASSISTments Longitudinal Data Mining Competition. This dataset collected data over a decade, which tracked students' intercations at the ASSISTments learning platform from middle school study in 2004-2007 to their high school course-taking, until they graduated from the college. Therefore, the average learning record of students in ASSISTments2017 is much longer than other dataset, which is beyond the length of 1,000. ASSISTments2017 also has rich side information, including but not limited to \textit{AveKnow} that indicates students' average knowledge level based on BKT, \textit{timeTaken} that represents the time spent on the current exercise, \textit{frIsHelpRequest} that represents whether the first response is a help request.
\end{itemize}

\subsubsection{Junyi Dataset}
The Junyi dataset \citep{ChangHC15} contains the problem log and exercise-related information on the Junyi Academy\footnote{http://www.junyiacademy.org/}, a Chinese e-learning website established in 2012 on the basis of the open-source code released by Khan Academy. In contrast to the ASSISTments datasets, Junyi has less exercsies and KCs, but includes an exercise hierarchy labeled by experts, the annotations of exercise relationship are also available. Therefore, many research works that focused on the knowledge structure in KT had utilized this dataset \citep{tong2020structure}. Junyi provides the prerequisite exercise of a specific exercise in the knowledge map, the topic and area of each exercise, as well as the coordiate position of the knowledge map.

\subsubsection{Eedi2020 Dataset}
The Eedi2020 dataset \citep{wang2020diagnostic} is also released in an achedemic challenge, i.e., NeurIPS 2020 Education Challenge\footnote{https://eedi.com/projects/neurips-education-challenge}. This dataset contains students’ answers to mathematics questions from Eedi, an online educational platform which millions of students interact with daily around the globe from school year 2018 to 2020. All exercises are multiple-choice problems with 4 possible answer choices, exactly one of which is correct. In Table \ref{data}, we give the statistics based on the training data in this competition, the total number of learning records in the full dataset exceeds 17 million. It worth noting that Eedi2020 gives students' exact answer choice so that we can also predict students options \citep{ghosh2021option}. Moreover, for the students, Eedi2020 records lots of valuable context information, including the \textit{Gender}, \textit{DateOfBirth}, \textit{PremiumPupil}. For the learning records, Eedi2020 also presents their \textit{Confidence}, \textit{GroupId}, \textit{QuizId}, and \textit{SchemeOfWorkId}. 

\subsubsection{Statics2011 Dataset}
Different from the above datasets that focus on mathematics exercises, the Statics2011 dataset \citep{2010statistics} is obtained from a college-level engineering statics course via online educational system developed by Carnegie Mellon University\footnote{https://pslcdatashop.web.cmu.edu/DatasetInfo?datasetId=507}. {The problems in college engineering course are quite complex, often comprising numerous independent sub-steps. Consequently, we treat each sub-problem as an exercise and calculate the number of exercises and KCs within this dataset.}

\subsubsection{EdNet Dataset}
The EdNet dataset \citep{ednet} is related to the english subject, which is consisted of students' learning records in the multi-platform AI tutoring system Santa in South Korea\footnote{https://github.com/riiid/ednet}. EdNet collected learning data of students over two years for their preparation of the e TOEIC (Test of English for International Communication) Listening and Reading Test. EdNet is now the largest public dataset in KT field with a total of 131,441,538 learning records from 784,309 students. Besides, it contains various features of students' learning actions, such as the specific learning material they have interacted, how much time they have spent for answering a given exercise. There are four differnet versions of EdNet, respectively named EdNet-KT1, EdNet-KT2, EdNet-KT3, and EdNet-KT4 with different extents. {We note that the students in this dataset may be in different learning states. Therefore we do not give the learning state information in Table \ref{data}.}
\begin{itemize}[leftmargin=*]
	\item{\textbf{EdNet-KT1.}}
	EdNet-KT1 contains students' basic exercise-answering logs. This dataset has 784,309 students, 13,169 exercises, 188 KCs, and a total of  95,293,926 learning records.
	Exercises in EdNet-KT1 are organized by bundles, i.e., a collection of exercises sharing a common passage, picture or listening material. Therefore, exercises come up in bundles and students have to answer all contained exercises when a bundle is given.
	
	\item{\textbf{EdNet-KT2.}}
	EdNet-KT2 recorded students' action sequences, which indicated their full learning behaviors. For example, a student who is not confident about the answer may alternately select among several answer choices before submitting. Such learning behaviors can reflect more fine-grained knowledge state of students. EdNet-KT2 contains three kinds of actions: \textit{enter} when student first receives and views a bundle , \textit{respond} when the student selects an answer choice to the exercise, and \textit{submit} when the student submits his final answers to the the given bundle. It is worth noting that EdNet-KT2 is a subset of EdNet-KT1.
	
	\item{\textbf{EdNet-KT3.}}
	On the basis of EdNet-KT2, EdNet-KT3 collected more students' learning activities, such as reading explanations or watching lectures. These learning activities have potential impacts on students' knowledge state so that they are valuable to be analyzed.
	
	\item{\textbf{EdNet-KT4.}}
	In EdNet-KT4, the very fine details of actions were provided. In particular, the following types of actions are added to EdNet-KT3: erase choice, undo erase choice, play audio, pause audio, play video, pause video, pay, refund, and enroll coupon.
\end{itemize}

\subsubsection{CodeWorkout Dataset}
The CodeWorkout dataset \citep{edwards2017codeworkout} is utilized in the 2nd Computer Science Educational Data Mining Challenge (CSEDM)\footnote{https://sites.google.com/ncsu.edu/csedm-dc-2021/home}. This dataset  is collected from a CS1 course in the Spring and Fall 2019 semesters at a public university in United States. It contains the code submissions from students for 50 coding problems, each requiring ~10-26 lines of code. {In this dataset, each exercise has a unique KC, thereby the number of KCs is equal to the number of exercises.} In total, there are 329 and 490 students in the Spring and Fall semesters who completed the course. Each dataset contains more than 65,000 code submissions, the scores of the submissions (\% of unit tests passes) are also available, as well as the compiler message if the compilation is not successful. The final grades of students are also provided for this dataset.

\subsection{Baselines}
The implementations of existing KT methods are not standardized, which may use different program languages (e.g., python, lua) and different deep learning frameworks (e.g., tensorflow, torch). Furthermore, some works did not well organize the codes systematically (e.g., the missing of running environments and dependencies), which brings difficulties in reproducing the models. To this end, we put forward the algorithm library for KT baselines, named EduKTM, which now has contained the concurrent popular works. EduKTM will be always under development for including the latest KT models, more algorithms and features are going to be added. Besides, we provide detailed guidelines for everyone who is interested in contributing to EduKTM.

It worth nothing that we do not provide the performance evaluation of these baselines using the aforementioned benchmark datasets. The reasons are two-folds: (1) For each baseline method, their experimental settings are quite different as they were designed to handle with various learning scenarios. No single KT model can always be the best one under various learning contexts. It is not fair to compare all baselines under a fixed setting.
(2) Existing evaluation standards  primarily concentrate on the performance of student performance prediction task, which cannot directly reflect the effectiveness of various KT models in practical applications.
Therefore, we decide to open source EduData and EduKTM. Researchers and practitioners can freely compare and select appropriate KT models based on their specific requirements in various application scenarios.

\section{Future Research Directions} \label{sec:future}
This survey has comprehensively reviewed the abundant current developments in the field of knowledge tracing, including fundamental KT models, their variations and typical applications. Despite this, as knowledge tracing is a relatively new but promising research area, there are still a substantial number of problems that require urgent resolution. In this section, we discuss several potential future research directions.

\subsection{Knowledge Tracing with Interpretability}

The performance of KT models is now evaluated indirectly through the student performance prediction task. The higher the precision of students' responses on future exercises, the better the KT model's performance. Nonetheless, interpretability plays a significant role in education, as students often express more concern about the 'why' than the 'what' of a learning decision \citep{li2023genetic}. Enhancing the interpretability of KT models is therefore crucial. Some educational theories, such as the Rasch model used in AKT \citep{CAKT} and the transfer of knowledge used in SKT \citep{tong2020structure}, could be considered for this purpose.  \citet{minn2022interpretable} noticed the significance of interpretable KT, particularly for deep learning models, as they have numerous parameters that are challenging to explain meaningfully. Thus, they \citep{minn2022interpretable} introduced a straightforward and interpretable KT approach, based on the causal relationships within the latent features extracted from students' behavioral data. 
\citet{10093075} attempted to introduce causal inference for explanatory analysis on KT, and they achieved more stable and explainable knowledge tracing based on the analysis results. 
It is imperative to further refine existing KT models or to explore additional methods for interpretable KT researches. This will lead to the production of more accurate and interpretable evaluations of students' knowledge states.

\subsection{Knowledge Tracing with Sparse Learning Interactions}
The acquisition of high-quality KT models necessarily requires a substantial amount of data to ensure training stability.  However, online learning systems often achieve less learning interactions in practical educational scenarios, thereby leading to the data sparsity problem. To address this issue, \citet{yang2021gikt} utilized graph convolutional network to include exercise-KC correlations. 
\citet{huang2023towards} turned to improve the attention mechanism.
More recently, researchers have employed contrastive learning to alleviate the data sparsity problem in KT \citep{song2022bi, dai2022contrastive}. For example, \citet{lee2022contrastive} presented a contrastive learning framework to enhance KT, which measure the semantical similarity between various learning interactions to effectively learn their representations. They further designed data augmentation methods to enhance the semantics of students' learning interactions. However, while the aforementioned methods alleviate the data sparsity problem in various aspects, there remains a need for further improvement in addressing this issue comprehensively.

\subsection{Knowledge Tracing with Subjective Exercises}
Most existing KT researches are capable of dealing with objective exercises, assuming student responses are binary. However, they overlook students' open-ended answers, often presented in natural language, on subjective exercises. As per the work by \citet{liu2022open}, an open-ended KT approach for computer science education was explored. This method aimed to predict students' precise solutions to programming questions, utilizing language models for assistance. 
Recent advancements in large language models offer promising potential to enhance the ability of KT models to comprehend the critical information regarding students' knowledge states embedded within their open-ended answers on subjective exercises \citep{kasneci2023chatgpt}.

\subsection{Knowledge Tracing with Students' Feedback}
Learning records are passive representations of students' knowledge states. In contrast, students' feedback provides us with their proactive understanding of their knowledge states, thereby offering direct and authentic indicators of their learning situation.  However, there are few KT models that take advantage of training data related to students' feedback, even though it can play an important role in fixing the KT results \citep{vasilyeva2007feedback}. \citet{2015Using} have noted that feedback plays a positive role in learning, which may promote transfer and retention in learning from worked-out examples. Therefore, incorporating students' feedback presents a meaningful avenue that could lead to significant improvements.

\subsection{Knowledge Tracing for General User Modeling}

Generally, user modeling refers to tools for characterizing users' behaviors (e.g., frequent locations), personal information (e.g., age, gender, and occupation) and latent features (e.g., interests and abilities), which facilitate the provision of targeted services for different users \citep{2001User}. As a type of latent feature modeling, knowledge tracing diagnoses the proficiency of users (not only individuals, but also  groups of individuals, like user teams and companies) on specific skills/concepts. Thus, in addition to education, knowledge tracing can be developed and applied across a wide range of domains for user modeling, including games, sports, and recruitment.

\subsection{Knowledge Tracing with Large Language Models}

{In recent years, the advancement of large language models (LLM), notably ChatGPT, has led to significant impacts and garnered considerable research interest worldwide \citep{zhao2023survey}. The use of LLM in education is promising to revolutionize the current learning pattern \citep{kasneci2023chatgpt}. The current study mainly focuses on generating learning materials, improving student-system interaction, and explaining educational contents. However, it remains  unclear how LLM can assist in understanding students' knowledge states. Given the powerful capabilities of LLM, it has the potential to enhance the generalization and interpretability of existing KT methods. Specifically, LLM can analyze the content of students' responses and evaluate their quality, which directly reflects students' knowledge states. Besides, LLM  can play the role of the instructor, answering questions for students and identifying their strengths and weaknesses. Furthermore, LLM itself can serve as a KT model, which can output reasonable results about students' knowledge states given their previous learning interactions.
Simultaneously, it is imperative to safeguard the privacy and security of student data when utilizing LLM \citep{pan2020privacy}. 
}

\section{Discussions and Conclusions} \label{sec:conclution}

In this survey, we conducted a thorough review of knowledge tracing. Specifically, we initially conducted a comprehensive review of existing fundamental KT models. Considering the complexity of online learning systems and the significant importance of KT research in practical scenarios, we investigated a broad range of variant KT research. Subsequently, we summarized some typical applications of KT in common educational scenarios. Furthermore, we released two algorithm libraries for KT-related datasets and baselines, thereby facilitating researchers and practitioners in selecting appropriate KT models based on their specific requirements. Finally, we outlined some potential future directions. 

We hope that this comprehensive survey of knowledge tracing will assist readers in understanding the problem of modeling students' dynamic knowledge states. It serves as a fundamental framework for both researchers and practitioners in future studies, fostering the development of KT.  The development of KT will directly benefit millions of students, with its impact potentially extending to a broader audience as online learning continues to evolve. In this context, KT will play an increasingly important role in enabling individuals to adapt to the ever-changing society \cite{wotto2020future}. Furthermore, the development of more refined KT methods tailored to students in various subjects and age groups will enhance their ability to comprehend students' individual knowledge states.  

However, we recognize some limitations of the current survey. 
Firstly, as new KT methods continue to emerge, although we have conducted a thorough investigation, there may be some representative works that we have overlooked. If necessary, we will incorporate these into our proposed framework in the future. 
Secondly, the complexity of KT methods warrants attention, as it continues to grow, particularly in those based on deep learning. The complexity of the model may enhance its accuracy, but it could also compromise its applicability, as users will question the reliability of complex models. Despite the fact that some studies have attempted to reveal the interpretability of complex KT models, for instance, by utilizing the xAI technique, there remains a significant distance to traverse before achieving a completely transparent deep learning KT model.

We posit that the application of KT methods in online education presents a promising avenue for research. It significantly enhances the learning experience for students while simultaneously alleviating the burden on teachers. Despite the numerous challenges and obstacles, researchers are be encouraged to overcome them and ensure a reliable and equitable access to students' evolving knowledge state in learning. 

\section*{Acknowledgment}

This research was supported by grants from the National Key Research and Development Program of China (Grant No. 2021YFF0901003), and the National Natural Science Foundation of China (Grant No. 62337001, U20A20229).

\bibliographystyle{IEEEtranN}
\begin{footnotesize}
	\bibliography{cite}
\end{footnotesize}

\vspace{-0.6cm}
\begin{IEEEbiography}[{\includegraphics[width=1in,height=1.25in,clip,keepaspectratio]{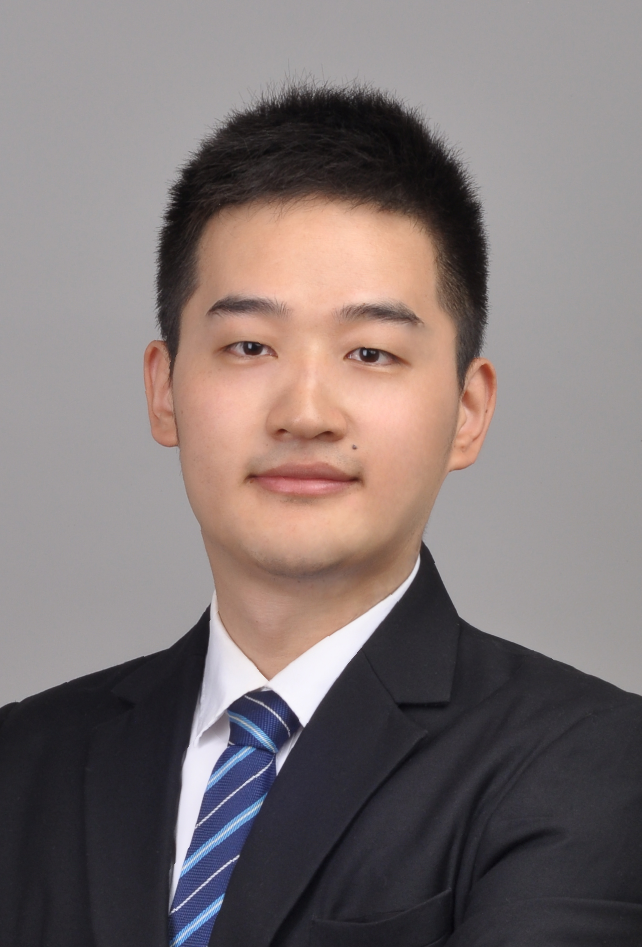}}]
	{Shuanghong Shen}  recieved the B.E. degree from Wuhan University, Wuhan, China, in 2018. He is currently working toward the Ph.D. degree in the School of Computer Science and Technology, University of Science and Technology of China. His main research interests include data mining, knowledge discovery, natural language processing and intelligent tutoring systems. He won the first prize in task 2 of the NeurIPS 2020 Education Challenge. He has published papers in referred conference proceedings, such as KDD2023, AAAI2022, SIGIR2022.
\end{IEEEbiography}
	
\vspace{-0.6cm}
\begin{IEEEbiography}[{\includegraphics[width=1in,height=1.25in,clip,keepaspectratio]{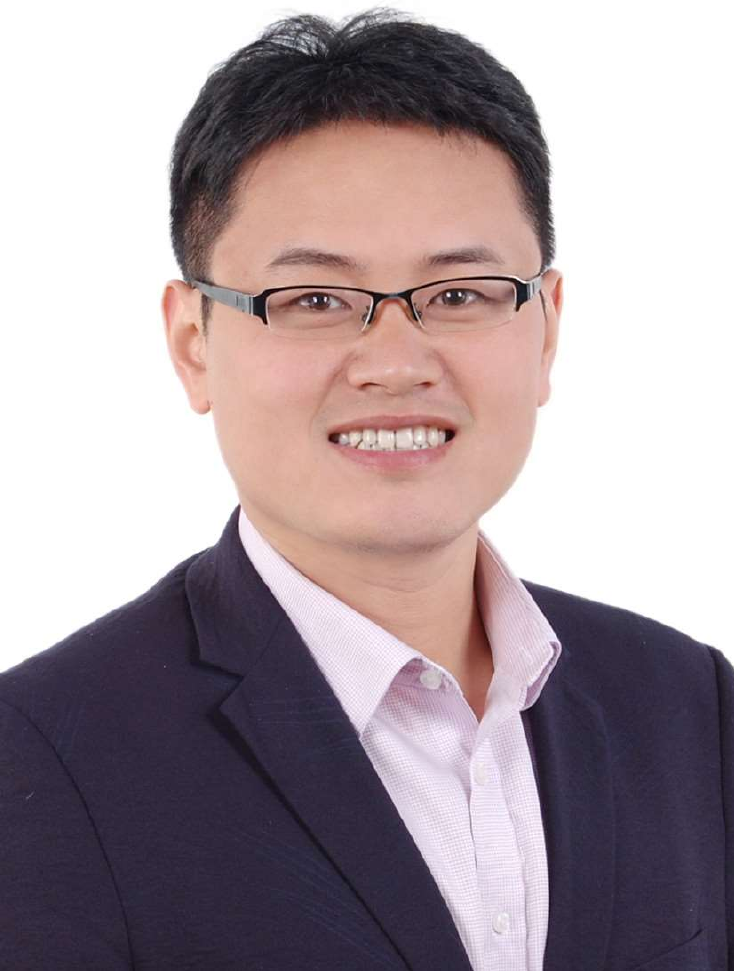}}]
	{Qi Liu} (Member, IEEE) received the Ph.D. degree from University of Science and Technology of China (USTC), Hefei, China, in 2013. He is currently a Professor in the School of Computer Science and Technology at USTC. His general area of research is data mining and intelligent education. He has published prolifically in refereed journals and conference proceedings (e.g., TKDE, TOIS, KDD). He is an Associate Editor of IEEE TBD and Neurocomputing. He was the recipient of KDD’ 18 Best Student Paper Award and ICDM’ 11 Best Research Paper Award. He is a member of the Alibaba DAMO Academy Young Fellow. He was also the recipient of China Outstanding Youth Science Foundation in 2019.
\end{IEEEbiography}

\vspace{-0.6cm}
\begin{IEEEbiography}[{\includegraphics[width=1in,height=1.25in,clip,keepaspectratio]{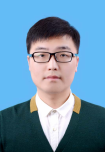}}]
	{Zhenya Huang} (Member, IEEE) received the B.E. degree from Shandong University, in 2014 and the Ph.D. degree from USTC, in 2020. He is currently an associate researcher of the School of Computer Science and Technology, University of Science and Technology of China (USTC). His main research interests include data mining, knowledge discovery representation learning and intelligent tutoring systems. He has published more than 30 papers in refereed journals and conference proceedings including TKDE, TOIS, KDD, AAAI. He has served regularly in the program committees of a number of conferences, and is reviewers for the leading academic journals.
\end{IEEEbiography}
\vspace{-0.6cm}

\begin{IEEEbiography}[{\includegraphics[width=1in,height=1.25in,clip,keepaspectratio]{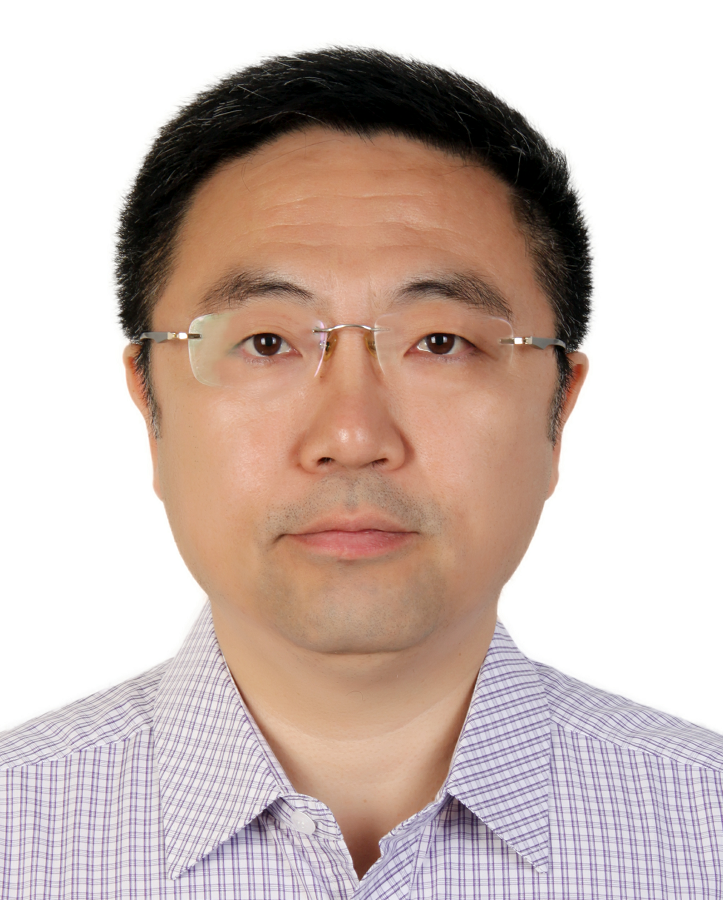}}]
	{Yonghe Zheng} is the funding dean of Research Institute of Science Education in Beijing Normal University, Professor in Education Faculty of BNU, PhD Supervisor. Prof. Zheng is the Vice president of the 8th Council of China Association of Children's Science Instructors and is the member of Teaching Guidance Committee of the Basic Education, the member of comprehensive group of the compulsory education curriculum revision committee, the science curriculum standard revision group; the General Secretary of the Awarding Scheme Council for outstanding computer teachers in Universities; the Vice director of the academic committee of Chinese society of Natural Science Museum; and former Director General of the Science Policy Bureau of National Natural Science Foundation of China. Prof. Zheng has engaged in project management of basic research and S\&T policy research for a long time. More than 60 articles have been published. And Zheng is the chief editor of Disciplinary and Interdisciplinary Science Education Research Journal. Prof. Zheng joined in Educational Faculty of Normal University in April, 2018, with specific research interests on science education and educational technology. Prof. Zheng established the Research Institute of Science Education of Beijing Normal University in November of 2019.
\end{IEEEbiography}
\vspace{-0.6cm}

\begin{IEEEbiography}[{\includegraphics[width=1in,height=1.25in,clip,keepaspectratio]{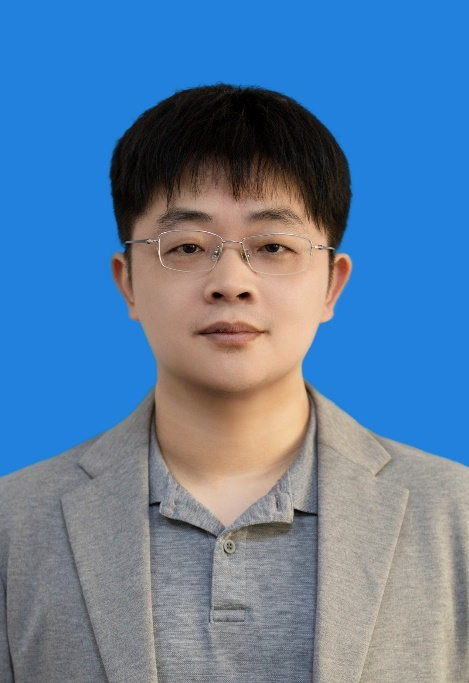}}]
	{Minghao Yin} (Member, IEEE) received the B.S. and M.S. degrees from Northeast Normal University, Changchun, China, in 2001 and 2004, respectively, and the Ph.D. degree from Jilin University, Changchun, China, in 2008, all in computer science. He is currently a professor, the dean of the School of Information Science and Information Technology at Northeast Normal University, and the director of the Key Laboratory of Intelligent Information Processing in Jilin Province Colleges and Universities. He also contributes as the Associate Editor for the IEEE Transactions on Learning Technologies (TLT). He has published multiple papers in prestigious journals and conferences, including Artificial Intelligence, IEEE Transactions, ACM Transactions, JAIR, AAAI, IJCAI, and WWW. His research interests include heuristic search, data mining, and combinatorial optimization.
\end{IEEEbiography}
\vspace{-0.6cm}

\begin{IEEEbiography}[{\includegraphics[width=1in,height=1.25in,clip,keepaspectratio]{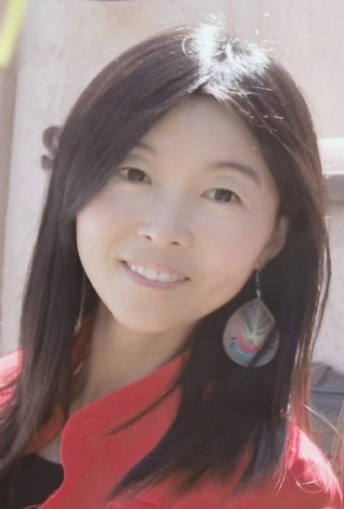}}]
	{Minjuan Wang} (Member, IEEE) obtained her B.A. in Chinese literature from the Beijing/Peking University in 1995, M.A. in comparative literature from Penn State University in 1997, and Ph.D. in information science and learning technologies from University of Missouri-Columbia in 2001. She is	a newly appointed Chair Professor of Emerging Technologies and Future Education at The Education University of Hong Kong. She has also been a professor and program chair of Learning design and technology at San Diego State University, California, USA since 2000. In addition, she serves as the Editor-in-Chief for the IEEE Transactions on Learning Technologies (TLT). Her research specialties are multidisciplinary, focusing on STEM education, new and emerging technologies in various educational settings, Metaverse and immersive learning, and	the design and implementation of artificial intelligence including AIGC for education and training. She is an internationally recognized scholar and has keynoted about 45 international conferences. She is recognized internationally for her research, publishing and dedicated service to IEEE and other scholarly communities. Dr. Wang is a member of IEEE and the IEEE Education Society. She also co-chairs the Education Society’s technical committee for immersive learning (TC-ILE). 
	
\end{IEEEbiography}
\vspace{-0.6cm}

\begin{IEEEbiography}[{\includegraphics[width=1in,height=1.25in,clip,keepaspectratio]{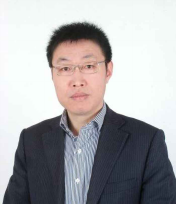}}]
	{Enhong Chen} (Fellow, IEEE) received the B.S. degree from Anhui University, Hefei, China, the M.S. degree from the Hefei University of Technology, Hefei, China, and the Ph.D. degree in computer science from the University of Science and Technology of China (USTC), Hefei, China, in 1989, 1992 and 1996 respectively. He is currently a Professor and the Executive Dean of School of Data Science, the Director of Anhui Province Key Laboratory of Big Data Analysis and Application. He has published a number of papers on refereed journals and conferences, such as TKDE, TIST, TMC, KDD, ICDM, NIPS and CIKM. His current research interests include data mining and machine learning, social network analysis, and recommender system. Dr. Chen was a recipient of the National Science Fund for Distinguished Young Scholars of China, the Best Application Paper Award on KDD-2008, the Best Student Paper Award on KDD-2018 (Research), and the Best Research Paper Award on ICDM-2011. He is an IEEE Fellow and an ACM Distinguished Scientist.
\end{IEEEbiography}
\vspace{-0.6cm}

\end{document}